\documentclass[twocolumn,showpacs,prc,unsortedaddress]{revtex4-1}

\usepackage{amssymb}
\usepackage[intlimits]{amsmath}
\usepackage{graphicx}
\usepackage{color}
\usepackage{times}
\usepackage[ansinew]{inputenc}
\usepackage[T1]{fontenc}
\usepackage[english]{babel}

\newcommand{\code}[1]{\textsc{#1}}
\newcommand{\incl}{\code{INCL4.5}}
\newcommand{\gemini}{\code{GEMINI}}
\newcommand{\geminipp}{\code{GEMINI++}}

\newcommand{\ud}{\mathrm{d}}
\newcommand{\etal}{et al.}

\def\clap#1{\hbox to 0pt{\hss#1\hss}}

\def\mathrlap{\mathpalette\mathrlapinternal}

\def\mathrlapinternal#1#2{%
\rlap{$\mathsurround=0pt#1{#2}$}}

\begin{document}

\title{Unified description of fission in fusion and spallation reactions}

\author{Davide Mancusi}
\affiliation{University of Li\`ege, AGO Department, all\'ee du 6 ao\^ut 17, b\^at. B5, B-4000 Li\`ege 1, Belgium}
\author{Robert J. Charity}
\affiliation{Department of Chemistry, Washington University, St.\ Louis, Missouri 63130, USA}
\author{Joseph Cugnon}
\affiliation{University of Li\`ege, AGO Department, all\'ee du 6 ao\^ut 17, b\^at. B5, B-4000 Li\`ege 1, Belgium}

\begin{abstract}
We present a  statistical-model description of fission, in the framework of compound-nucleus decay, 
which is found to simultaneously 
reproduce data from both heavy-ion-induced fusion reactions and proton-induced spallation reactions at around 1 GeV.
For the spallation reactions, the initial compound-nucleus population is predicted by the  Li\`{e}ge Intranuclear Cascade Model.
We are able to reproduce experimental fission probabilities and fission-fragment mass distributions 
in both reactions types with the same parameter sets. However, no unique parameter set was obtained for the fission probability.
The introduction of fission transients can be offset by an increase of the ratio of level-density parameters for the saddle-point
and ground-state configurations. Changes to the finite-range fission barriers could be offset by a scaling of the Bohr-Wheeler
decay width as predicted by Kramers. The parameter sets presented allow accurate prediction of fission probabilities
for excitation energies up to 300~MeV and spins up to 60~$\hbar$. 
\end{abstract}

\maketitle

\section{Introduction}

Although seventy years have passed since the seminal works of Bohr and Wheeler \cite{bohr-fission} and Weisskopf and Ewing
\cite{weisskopf-evaporation} and the establishment of a qualitative understanding of the de-excitation mechanism of
excited nuclei, quantitatively accurate and universally applicable models do not yet exist. This is partly due to the
vastness of the amount of nuclear data that must be fed into the models, and partly to the uncertainties in the
fundamental ingredients of de-excitation, such as level densities and emission barriers. Even the choice of the
mathematical formalism, however, is not devoid of confusion, as it was already pointed out by Moretto
\cite{moretto-binarydecay} and Swiatecki \cite{swiatecki-binarydecay}.

One  way to lift the degeneracy of the ingredients of the model is to explore diverse regions of the compound-nucleus 
parameter space. A systematic study of nuclei with different masses, excitation energies, spins and isospins would
be sensitive to most of the assumptions of the de-excitation model. The long-term goal of such an investigation would be
to identify a minimal set of physical ingredients necessary for a unified quantitative description of nuclear
de-excitation chains.

The production of excited compound nuclei can proceed from several entrance reactions.
There has been a long history of compound-nucleus studies using heavy-ion-induced fusion reactions. 
These reactions allow one to specify the compound-nucleus mass, charge and excitation energy;
however, a distribution of compound-nucleus spins is obtained. Statistical-model parameters such as fission barriers
are quite sensitive to spin.
Heavy-ion-induced  fission probabilities, evaporation spectra, residue masses can generally be reproduced
in statistical-model calculations. However, some fine tuning of the statistical-model parameters to the mass region or 
reaction is often needed. A unified description over all mass regions is still lacking even for these reactions. 
Some work towards this, concentrating of the parameters describing the shape of the evaporation spectra, is presented 
in Ref.~\cite{Charity10}.   

Another typical entrance channel for the production of excited compound nuclei is spallation. The present interest in
spallation derives mainly from the applications to Accelerator-Driven Systems (ADS), namely accelerator-based reactors for
the transmutation of nuclear waste. At incident energies relevant for transmutation (a few hundred MeV), an appropriate
theoretical tool for the description of proton-nucleus reactions is the coupling of an intranuclear-cascade model (INC)
with a nuclear de-excitation model. It is assumed in the INC framework that the incoming particle starts an avalanche of
binary collisions with and between the target nucleons. When the cascade stage ends, an excited and thermalized remnant is
formed, with a basically unchanged density. In the subsequent de-excitation stage, the
remnant gets rid of the excess energy by particle evaporation and/or fission.
For these reactions, the need for a model to predict the initial compound-nucleus mass, charge, excitation, and spin 
distributions adds some uncertainty in our ability to constrain the statistical-model parameters by fitting data.  
However, spallation reactions allow us to explore different regions of compound-nucleus spin and excitation energy 
than can be probed with fusion reactions alone and thus can be important in parameter fitting.

The role of a transient fission width is currently of some controversy. Fission transients are where the fission decay 
width is not constant, but increases from zero towards its equilibrium value \cite{Grange83}.
Fission transients were first introduced to
help explain the large number of neutrons emitted from a fissioning system before the scission point was 
reached \cite{hilscher-prescissionN}. The statistical model assumes there is an equilibrium is all degrees of freedom including the 
deformation degrees of freedom associated with fission. If all compound nuclei have spherical shapes initially, then they
 cannot instantaneously fission as it takes as finite time to diffuse towards the saddle and subsequently the scission point.
The transient time, the time scale necessary for the system to explore large fluctuations in the deformation degrees of freedom, is a function of the  viscosity of the nucleus. The predicted fission probability is also very sensitive to the assumed initial deformation \cite{Charity04}
which may depend on the entrance channel.

The transient time is often called a fission delay as fission is suppressed during this period. If the excitation energy 
of the compound nucleus is large enough, then there will be a probability of emitting a light or possibility even an 
intermediate-mass fragment during the fission delay. Neutron emission lowers the excitation energy and 
charged-particle emission also lowers the fissility of the nucleus by increasing its fission barrier. 
These effects will lead to a reduced fission probability after the fission delay is over. An experimental test of this 
idea would be the observation of reduced fission probability or enhanced evaporation-residue survival at high excitation energies 
which cannot be explained in terms of the statistical model. At present there is some controversy over the need for fission transients.

A number of theoretical  studies \cite{Frobrich93} reproduce experimental 
fission probabilities and pre-scission neutron multiplicities with transient fission widths 
when the  viscosity increases with the mass of the compound nucleus.
Transient fission has also been invoked to explain the relatively
large number of evaporation residues measured for the very fissile
$^{216}$Th compound nuclei formed in $^{32}$S+$^{184}$W reactions,
as compared to a statistical-model prediction \cite{Back99}. 
Alternatively other studies have reproduced fission probabilities with no transient effects \cite{Moretto95} 
and Lestone and McCalla  \cite{Lestone09} suggest that fission transients are 
unimportant whenever the nuclear temperature is less than the fission 
barrier.
Similarly, in very high excitation-energy data obtained in 2.5-GeV proton-induced reactions, 
no transients were needed in reproducing the measured fission yields \cite{Tishchenko05}.

This paper discusses the application of the \geminipp\ de-excitation model \cite{charity-gemini++} to the description of
fission in fusion and spallation reactions. In the latter case, the description of the entrance channel is provided by a
coupling to the Li\`ege Intranuclear Cascade model (\code{INCL}) \cite{boudard-incl}. Both \code{INCL} and \geminipp\ are
among the most sophisticated models in their own fields. The present work also represents the first thorough discussion of
their coupling.

We compare the predictions of the models with experimental residue yields in spallation studies and with fission and
evaporation-residue excitation functions measured in heavy-ion induced fusion reactions. The choice of the observables was
motivated by considerations about their sensitivity to fission and by the availability of experimental data. An exhaustive
discussion should of course take into account other observables (e.g.\ double-differential particle spectra) and the
competition of fission with the other de-excitation channels, but this is outside the scope of the present paper. Therefore, we
describe how the parameters of the statistical-decay model have been consistently adjusted to reproduce the data and
discuss to what extent a successful unified description of these reactions has been achieved. Finally, we explore whether
all the data can be described within the statistical model or whether transient fission decay widths are needed.

\section{The Models}

We shall now turn to the description of the most important features of the models we have considered. The codes will not
be analyzed in detail, but only the most important features will be outlined.

\subsection{\geminipp}\label{sc:geminipp}
\label{Sec:geminipp}
\geminipp\ is an improved version of the \gemini\ statistical decay model, developed by R.~J. Charity \cite{charity-gemini} with the goal of
describing complex-fragment formation in heavy-ion fusion experiments. The de-excitation of the compound nucleus proceeds through a
sequence of binary decays until particle emission becomes energetically forbidden or improbable due to competition with
gamma-ray emission.

Since compound nuclei created in fusion reactions are typically characterized by large intrinsic angular momenta, the
\gemini\ and \geminipp\ models explicitly consider the influence of spin and orbital angular momentum on particle
emission. Moreover, \gemini/\geminipp\ do not restrict binary-decay modes to nucleon and light-nucleus evaporation, which
are the dominant decay channels, but allow the decaying nucleus to emit a fragment of any mass. The introduction of a
generic binary-decay mode is necessary for the description of complex-fragment formation and is one of the features that
set \gemini/\geminipp\ apart from most of the other de-excitation models.

Emission of nucleons and light nuclei ($Z\leq{}$2,
3 or 4, depending on the user's choice) is described by the Hauser-Feshbach evaporation formalism
\cite{hauser-evaporation}, which explicitly treats and conserves angular momentum.  The production of heavier fragments is
described by Moretto's binary-decay formalism  \cite{moretto-binarydecay}.
However for symmetric divisions of heavy compound nuclei,
the Moretto formalism employing Sierk's  Finite-Range calculations \cite{sierk-asyBar,carjan-asyBar} 
fails to reproduce the mass distribution of decay products (Sec.~\ref{sec:fmass}). However for light systems, the Moretto formalism 
works quite well \cite{sierk-asyBar, charity-gemini,han-Br} 
and is still used in \geminipp.
Also for the heavier systems, but for mass-asymmetries outside of the 
symmetric fission peak, the Moretto formalism is still used. 
Otherwise, the total fission yield 
is obtained from the Bohr-Wheeler formalism \cite{bohr-fission} and the width of the fission-fragment 
mass distribution is taken 
from systematics compiled by Rusanov \etal\ \cite{rusanov-mass} 
(see Sec.~\ref{sec:fmass}).

Table~\ref{tb:geminipp_processes} summarizes the de-excitation mechanisms featured by \geminipp.

\begin{table}
  \caption{\label{tb:geminipp_processes}
    List of de-excitation processes featured by the \geminipp\ model. The symbol $Z$ represents the charge number of
    the emitted particle and $Z_\text{switch}$ can be chosen to be 2, 3 or 4.}
  \begin{ruledtabular}
    \begin{tabular}{c|c|c}
      \textbf{Process}&\textbf{Model}&\textbf{Notes}\\
      \hline
      \parbox[][][c]{0.3\linewidth}{\smallskip{}\centering light-particle evaporation\smallskip{}}&
      \parbox[][][c]{0.4\linewidth}{\smallskip{}\centering Hauser-Feshbach \cite{hauser-evaporation}\smallskip{}} &
      \parbox[][][c]{0.2\linewidth}{\smallskip{}\centering $Z\leq Z_\text{switch}$\smallskip{}}\\

    \parbox[][][c]{0.3\linewidth}{\smallskip{}\centering binary decay\smallskip{}}
    & \parbox[][][c]{0.4\linewidth}{\smallskip{}\centering Moretto \cite{moretto-binarydecay}\smallskip{}}
    & \parbox[][][c]{0.2\linewidth}{\smallskip{}\centering$Z>Z_\text{switch}$\smallskip{}}\\

    \parbox[][][c]{0.3\linewidth}{\smallskip{}\centering fission\smallskip{}} &
    \parbox[][][c]{0.4\linewidth}{\smallskip{}\centering Bohr-Wheeler \cite{bohr-fission}\smallskip{}}
    & \parbox[][][c]{0.2\linewidth}{\smallskip{}\centering only in heavy systems\smallskip{}}\\

    \parbox[][][c]{0.3\linewidth}{\smallskip{}\centering partition in fission\smallskip{}} &
    \parbox[][][c]{0.4\linewidth}{\smallskip{}\centering Rusanov \etal\ \cite{rusanov-mass} \smallskip{}}
    & \parbox[][][c]{0.2\linewidth}{\smallskip{}\centering\smallskip{}}
  \end{tabular}
\end{ruledtabular}
\end{table}

The parameters of the model associated with evaporation have been adjusted to reproduce data 
from heavy-ion-induced fusion reactions. This is described in Ref.~\cite{Charity10} in more detail, but we briefly
list
the important adjustments for that work . 
In order to fit experimental light-particle kinetic-energy spectra, the transmission coefficients in the 
Hauser-Feshbach formalism were calculated for a distribution of Coulomb barriers associated with thermal fluctuations. 
The nature of fluctuations is not entirely clear, they may be fluctuations in compound-nucleus shape and/or its density 
and/or its surface diffuseness. 

Level densities were calculated with the Fermi-gas form:
\begin{equation}
\rho(E^*,J) \sim \exp \left( 2 \sqrt{ a(U) U}\right)
\label{eq:ld}
\end{equation}
where $E^*$ is the total excitation energy, $J$ is the spin and $U$ is the thermal excitation energy after the pairing, rotational, and deformation energies are removed.
The level-density parameter used should be considered an effective value as no collective-enhancement 
factors are used in the level density formula of Eq.~(\ref{eq:ld}).

The level-density parameter $a(U)$ is excitation-energy dependent with an initial fast dependence due to the 
washing out of shell
effects following Ref.~\cite{Ignatyuk75}  and a slower dependence needed to fit the evaporation spectra. The 
shell-smoothed level-density parameter was assumed to have the form
\begin{equation}
\widetilde{a}\left( U\right) =\frac{A}{k_{\infty }-(k_{\infty }-k_{0})\exp
\left( -\frac{\kappa }{k_{\infty }-k_{0}}\frac{U}{A}\right) }
\label{eq:fita}
\end{equation}%
which varies from $A/k_0$ at low excitation energies to $A/k_{\infty}$ at large values. Here $k_0$=7.3~MeV, consistent 
with neutron-resonance counting data at excitation-energies near the neutron separation
energy, and $k_{\infty}$=12~MeV. The parameter $\kappa$ defines the rate of change of $\widetilde{a}$ 
with energy and it is essentially zero for 
light nuclei (i.e.\ a constant $\widetilde{a}$ value) and increases roughly exponentially with $A$ 
for heavier nuclei.
Although we expect a decrease in the level-density parameter with $U$ due to decreasing 
importance of long-range correlations with increasing excitation energy 
(due to washing out of collective-enhancement factors and also the reduction of the 
intrinsic level-density parameter), the strong mass dependence cannot be explained at present.

The strong excitation-energy dependence of $\widetilde{a}$ for heavy nuclei leads to increased 
nuclear temperatures which 
enhance very weak decay channels. For very fissile systems, these weak decay channels 
include \textit{n}, \textit{p}, and $\alpha$ evaporation and thus \geminipp\ calculations predict enhanced 
evaporation-residue production consistent with some experimental data. 
These enhanced evaporation-residue yields had previously been interpreted as a consequence of transient fission 
\cite{Back99}.  Clearly the excitation-energy dependence of $\widetilde{a}$ is very important in understanding the 
role of transient fission. However we note that for fission, the dominant decay mode in fissile nuclei, the yield is 
decreased relatively little by the increased temperature.    
 
\subsection{\incl}

The \code{INCL} model \cite{boudard-incl} can be applied to collisions between nuclei and pions, nucleons or light nuclei of
energy lower than a few GeV. The particle-nucleus collision is modelled as a sequence of binary collisions among the
particles present in the system; particles that are unstable over the time scale of the collision, notably $\Delta$
resonances, are allowed to decay. The nucleus is represented by a square potential well whose radius depends on the
nucleon momentum; thus, nucleons move on straight lines until they undergo a collision with another nucleon or until they
reach the surface, where they escape if their total energy is positive and they manage to penetrate the Coulomb
barrier.

The latest version of the \code{INCL} model (\incl) includes, among other things, isospin- and energy-dependent nucleon
potentials, an isospin-dependent pion potential and a new dynamical coalescence algorithm for the production of light
clusters (up to $A=8$ with the default program options). A comprehensive description will be published in the near future
\cite{cugnon-preparation}.

The \code{INCL} model simulates a complete cascade event, its output being the velocities of all the emitted
particles. The characteristics of the remnant (its mass, charge, momentum, excitation energy and intrinsic angular
momentum) are derived from the application of conservation laws and are passed to the chosen de-excitation code; the
latter simulates the decay of the remnant into a nuclear-stable residue plus a number of nucleons, nuclei and/or gamma
rays.

The \incl\ model is not to be considered as an adjustable model. It does contain parameters, but they are either taken
from known phenomenology (such as the matter density radius of the nuclei) or have been adjusted once for all (such as the
parameters of the Pauli blocking or those who determine the coalescence module for the production of the light charged
clusters). Therefore adjusting \incl/\geminipp\ on the experimental data basically amounts to the adjustment of the
\geminipp\ parameters. One should keep in mind that \incl\ brings in its own physics features and limitations.  For our
purpose here, they essentially determine the distributions of the remnant properties. These quantities cannot be compared
directly with experimental data, but the predictions of \incl\ concerning those observables that can be confronted
directly to experiment, namely the high energy parts of particle spectra, are of rather good quality, as it was shown
recently \cite{cugnon-incl45_nd2010}.

\section{Adjustment of Fission Yields}\label{sec:adjustment}

The assumption of thermal equilibrium implied by the statistical-decay hypothesis implies that the excited nucleus cannot
keep any memory of the entrance channel. One of the main aspirations of the \geminipp\ development is to provide a unified
and coherent description of nuclear de-excitation in spallation and fusion reactions at the same time.

\begin{figure}[tbp]
\includegraphics*[scale=0.4]{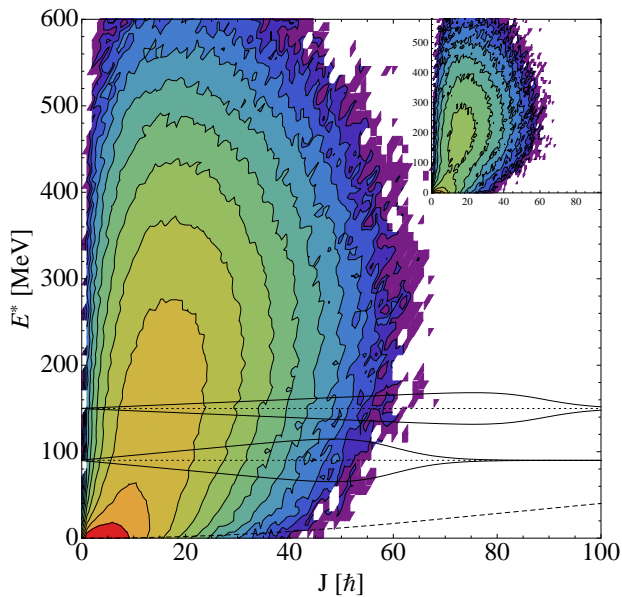}
\caption{(Color online) Comparison of the distributions of excitation energy and spin populated in the
  $^{19}$F+$^{181}$Ta$\rightarrow^{200}$Pb fusion reaction for $E^*={}$90,
  150~MeV (horizontal lines) with the \incl\ prediction for the 1-GeV \textit{p}+$^{208}$Pb spallation reaction
  (contours, logarithmically spaced). The inset shows the distribution of fissioning remnants for the spallation reaction
  (same contour levels). The dashed line is the macroscopic yrast line from Sierk \protect\cite{Sierk86}. }
\label{fig:ejmap}
\end{figure}

The degrees of freedom in the model induce different characteristic dependencies of the fission width on the remnant spin
and excitation energy, because fission is at the very least sensitive to spin, the fade-out of shell and 
collective effects, level densities and fission barriers. Since variations in some of the free parameters can produce similar
effects, it is difficult to disentangle the various contributions and to put stringent constraints on the de-excitation
model just by looking at experiments of a single type. However, fusion and spallation reactions populate different regions
of the compound-nucleus spin/excitation-energy plane. A comparison of the populations in the $E^*$-$J$ plane is shown in
Fig.~\ref{fig:ejmap} for the 1-GeV \textit{p}+$^{208}$Pb spallation reaction and the
$^{19}$F+$^{181}$Ta$\rightarrow^{200}$Pb fusion reaction.  The spallation population is represented by the
contours and two examples of the fusion distributions for $E^*$=90, 150~MeV are shown by the thick horizontal shapes, the
thickness of which is proportional to the population. The inset shows the distributions of remnants leading to fission in
the case of spallation; the contour levels are the same as in the main plot. As a guide for $^{200}$Pb, the macroscopic
yrast line from Sierk \cite{Sierk86} is indicated.

For the spallation reaction, the \incl\ model predicts average
values of about 167~MeV and 16.5~$\hbar$, but both distributions are quite broad and extend up to $\sim{}$650~MeV
and $\sim{}$50~$\hbar$,
respectively. On the other hand, the fusion reactions we considered are characterized by higher spins and 
lower excitation energies.
We concentrate only on complete-fusion reactions, where the excitation energy of the compound nucleus is defined
entirely from energy conservation. The requirement of complete fusion restricts us to projectile bombarding
energies of less then 10 MeV/$A$, where incomplete fusion and pre-equilibrium process are small.
We can explore somewhat higher 
excitation energies with more symmetric entrance channels, but high spins will still be populated.
 Thus it is clear that the comparison between spallation and fusion data represents a
promising tool to extend the predictive power of the model over a wide region of mass, energy and spin.

The outlined simultaneous-fitting strategy would be meaningless if the two types of data probed separate model-parameter
subsets; separate fits would then be just as effective as a combined fit. This is not the case, as Fig.~\ref{fig:ejmap}
suggests. Spallation is characterised by a large spread in the compound-nucleus excitation energy; however, the
excitation-energy dependence of fission can be tested in fusion reactions, too, using excitation curves. Thus, both types
of data sets are probably sensitive to the model parameters connected with the excitation-energy dependence of fission. A
similar argument can be produced for the model parameters connected with the spin dependence. The model parameters that
are sensitive to the mass and charge of the fissioning nucleus are probed by spallation through the distribution of
cascade remnants, and by fusion when one considers neighbouring compound nuclei. We then conclude that the fusion and
spallation probe different, but overlapping areas of the model-parameter space. Under these conditions, simultaneous
fitting is most likely advantageous.

The fusion reactions used in this study are listed in 
Table~\ref{Tbl:fus} with the range of excitation energies probed and the appropriate references for the data. 
In most cases we have selected data where both the evaporation-residue and fission cross sections have been determined. 
The sum of these two quantities gives the total fusion cross section and this is used to constrain the compound-nucleus 
spin distribution. We assume the spin distribution has a roughly triangular shape characterized by a maximum value $J_0$ 
with a smooth cutoff characterized by the parameter $\delta J$, i.e.\ 
\[
\sigma_\text{fus}(J) = \pi \lambdabar^2 (2J+1) \frac{1}{1+\exp \left( \frac{J-J_0}{\delta J}\right)}.
\]
The parameter $J_0$ is determined from the total fusion cross section
\[
\sigma_\text{fus} = \sum_{J=0}^{\infty} \sigma_\text{fus}(J)
\]
and $\delta J$ is set to values from 3 to 10~$\hbar$, with the larger values associated with the heavier projectiles. These values are similar to estimates obtained in Refs.~\cite{Charity86,Lesko86,Ackermann97}  

Generally the fission cross section is only sensitive to the value of $\delta J$ at excitation energies where the fission probability is small and rises rapidly with $J$ \cite{Charity86}. 
The values of $\delta J$ assumed in the following
calculations are also listed in Table~\ref{Tbl:fus}.
   
 \begin{table}
\caption{\label{Tbl:fus}Experimental fission and evaporation-residue data used in this work.}
\begin{ruledtabular}
\begin{tabular}{cccccc}
CN   &  reaction & $E^*$ range & $\sigma_\text{ER}$ & $\sigma_\text{fus}$ & $\delta J$ \\
     &           &  [MeV] &             &                & [$\hbar$]  \\
\hline
$^{156}$Er &  $^{64}$Ni + $^{90}$Zr    &  25-82 & \cite{Janssens86} & \cite{Wolfs89} & 10 \\
$^{158}$Dy &  $^{19}$F + $^{139}$La  &  68-94 & \cite{Charity86}  & \cite{Charity86} & 2.3 \\
$^{160}$Yb &  $^{60}$Ni + $^{100}$Mo &  95-249 & \cite{Charity03}  &                  & 10 \\
$^{168}$Yb &  $^{18}$O + $^{150}$Sm  &  63-126 & \cite{Charity03,Charity03a} & \cite{Charity03,Charity03a} & 4 \\
$^{178}$W  &  $^{19}$F+ $^{159}$W    &  54-95  & \cite{Charity03}  & \cite{Charity03} &4.2 \\
$^{188}$Pt &  $^{19}$F + $^{169}$Tm  &  50-91  & \cite{Charity03}  & \cite{Charity03} & 4.4 \\
$^{193}$Tl &  $^{28}$Si + $^{165}$Ho &  65-249 & \cite{Fineman94} &                  & 10 \\
$^{200}$Pb &  $^{19}$F + $^{181}$Ta  &  49-153 & \cite{Hinde82,Caraley00} & \cite{Hinde82,Caraley00} & 4.7 \\ 
$^{200}$Pb &  $^{30}$Si + $^{170}$Er &  48-84 & \cite{Hinde82}    & \cite{Hinde82}   &  10 \\
$^{216}$Th &  $^{32}$S + $^{184}$W   &  125-203 & \cite{Back99}     & \cite{Keller87}  & 10 \\
$^{216}$Ra &  $^{19}$F + $^{197}$Au  &  39-86 & \cite{Berriman01} & \cite{Berriman01} & 3 \\
$^{216}$Ra &  $^{30}$Si + $^{186}$W  &  39-83  & \cite{Berriman01} & \cite{Berriman01}   & 10 \\
$^{224}$Th &  $^{16}$O + $^{208}$Pb  &  26-83 & \cite{Brinkmann94} & \cite{Videbaek77,Back85} & 4 \\
\end{tabular}
\end{ruledtabular}
\end{table}

For the $^{28}$Si+$^{165}$Ho and $^{60}$Ni+$^{100}$Mo reactions, only evaporation residue data has been measured. 
However for these
reaction, the $J_0$ values associated with fusion-like reactions are very large and the higher $J$ values all
go into fission. The evaporation residue yield is therefore not sensitive to $J_0$ and is entirely determined by fission
competition at the lower $J$ values. Blann \etal\ \cite{Blann82} termed this a saturation analysis as the 
higher $J$ values are saturated by fission.

For spallation reactions, we focused our efforts on proton-induced fission reactions on $^{197}$Au
\cite{benlliure-gold}, $^{208}$Pb \cite{enqvist-lead} and $^{238}$U \cite{bernas-u_fission} at 1 GeV, measured
in inverse kinematics with the FRagment Separator (FRS) at SIS, GSI, Darmstadt, Germany. An additional experimental
data-set for \textit{p}+$^{208}$Pb at 500 MeV exists \cite{fernandez-lead}, but new measurements seem to indicate
that the fission cross section was overestimated by about a factor of 2 \cite{schmidt-nd2007}. We decided to normalize
Fernandez \etal's total fission cross section to the cross section measured by the CHARMS collaboration, assuming that the
fission distribution had been correctly measured. The reader should nevertheless keep in mind the normalization
uncertainties associated with this data set.

The simultaneous-fitting strategy, however, cannot be applied to pre-scission neutron multiplicities, since there are no
such data (to our knowledge) for spallation reactions. Moreover, accurate modelling of the pre- and post-scission neutron
data is likely to introduce new ingredients and parameters related to the the saddle-to-scission descent (e.g. the
viscosity of the saddle-to-scission motion). It is not obvious whether the extra constraint provided by the new data would
overcompensate the increase in the number of degree of freedom and lead to an effective decrease of the model uncertainty,
all the most so in the lack of relevant spallation data. Thus, we will not discuss pre-scission neutron multiplicities in
this work.

\subsection{Modifications of the Fission Width}

The Bohr-Wheeler fission width,
\[
\Gamma_\text{BW}=\frac{1}{2\pi\rho_\text{n}(E^*,J)}\int\ud\epsilon  \, \rho_\text{f}(E^*-B(J)-\epsilon,J)\text,
\]
is sensitive to the choice of the fission barrier $B$ and to the level-density parameters $a_\text{f}$ and $a_\text{n}$
associated with the saddle-point and ground-state configurations. The $U$ dependence of the level-density parameter was
initially assumed to be identical for the ground state and the saddle point, and it is described by
Eq.~\eqref{eq:fita}. However, the saddle-point level-density parameter $a_\text{f}$ was scaled by a constant factor with
respect to the corresponding ground-state level-density parameter $a_\text{n}$, to account for the increased surface area
of the saddle-point configuration \cite{Toke81}. In what follows, we will refer to the scaling factor as ``the
$a_\text{f}/a_\text{n}$ ratio'', for simplicity.

A number of modifications to the Bohr-Wheeler width have been proposed. 
In a one-dimension derivation of the escape rate over a 
parabolic barrier for high viscosity, Kramers \cite{Kramers40} obtained 

\begin{equation}
\Gamma_K = \left[\sqrt{1+\left(\frac{\gamma}{2\omega}\right)^2} - \frac{\gamma}{\omega}\right]  \Gamma_\text{BW}
\label{eq:Kramers}
\end{equation}
where $\gamma$ is the magnitude of the viscosity, $\omega$ is frequency associated with the inverted 
parabolic barrier, and the factor scaling the Bohr-Wheeler decay width is less than unity.
Now $\omega$ is not expected to be a strong function of mass or spin, 
and if $\gamma$ is also constant, then the Kramers and the Bohr-Wheeler values differ by approximately  
a constant scaling factor. For this reason we have allowed a constant scaling to the Bohr-Wheeler width.

Lestone \cite{lestone-tilting} developed a treatment of fission which explicitly included the tilting collective degree of freedom 
at saddle point. Tilting is where the compound nucleus's spin is not perpendicular to the symmetry axis. 
For strongly-deformed objects like the saddle-point configuration, this costs energy and thus decreases the fission probability. 
The decay width becomes  
\[
\Gamma_\text{Lestone} = \Gamma_\text{BW} \frac{\sum_{K=-J}^{J} \exp \left( -\frac{K^2}{2 I_\text{eff}}\right)}{2 J + 1}
\]
where the summation is over $K$, the projection of the spin on the symmetry axis and 
\[
\frac{1}{I_\text{eff}} = \frac{1}{I_\parallel} - \frac{1}{I_\perp}
\]
and $I_{\parallel}$ and $I_{\perp}$ are the saddle-point moments of inertia parallel and perpendicular to 
the symmetry axis, respectively. In this work, the moments of inertia as well as the spin-dependent saddle-point energies were
taken from the finite-range calculations of Sierk \cite{Sierk86}. Deviations from the Bohr-Wheeler value are largest for 
the highest spins and thus the Lestone modification will be more important in fusion reactions.

We have tried to reproduce simultaneously fission cross sections
from fusion and spallation experiments by:
\begin{itemize}
\item Adding a constant to the Sierk fission barriers for all spins. 
\item Scaling the decay width by a constant factor.
\item Adjusting the $a_\text{f}/a_\text{n}$ ratio.
\item Using either the Bohr-Wheeler or the Lestone formalism.
\item Introducing a constant fission delay.
\end{itemize}

\subsection{Fission Probability}
\label{sec:fprob}

\begin{figure}[tbp]
\includegraphics*[scale=0.4]{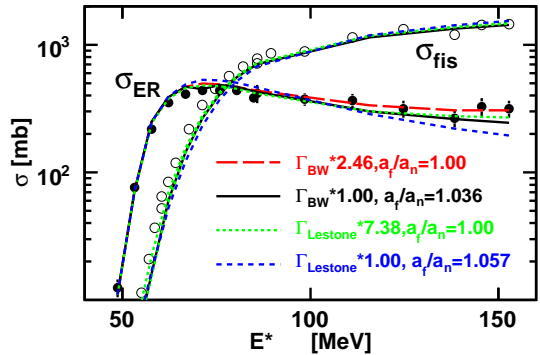}
\caption{(Color online) Comparison of \geminipp\ predictions to the experimental evaporation-residue 
and fission excitation functions for the $^{19}$F+$^{181}$Ta reaction.}
\label{fig:fitPb200}
\end{figure}

Examples of fits to the $^{19}$F+$^{181}$Ta$\rightarrow^{200}$Pb fission and evaporation-residue
excitation functions are shown in Fig.~\ref{fig:fitPb200}. As the sum of these quantities (the fusion cross section) is 
fixed in the calculations, the degree to which the fission probability is reproduced is best 
gauged by the fit to the smaller quantity, i.e.\ 
 $\sigma_\text{fis}$ at low bombarding energies  and $\sigma_\text{ER}$ at the higher values. Good fits were obtained with 
 $\Gamma_\text{BW}\times 2.46, a_\text{f}/a_\text{n}=1.00$ (long-dashed curves), 
$\Gamma_\text{BW}\times 1.00, a_\text{f}/a_\text{n}=1.036$ 
(solid curves),   $\Gamma_\text{Lestone}\times 7.38, a_\text{f}/a_\text{n}=1.00$ (dotted curves), and 
$\Gamma_\text{Lestone}\times 1.00, a_\text{f}/a_\text{n}=1.057$ (short-dashed curves). 
 The $\Gamma_\text{BW}\times 2.46, a_\text{f}/a_\text{n}=1.00$ calculation is also almost identical to a  
$\Gamma_\text{BW}\times 1.00, a_\text{f}/a_\text{n}=1.00$ calculation (not plotted) obtained with the Sierk fission barrier reduced by 1.0 MeV. With an even 
larger barrier reduction factor, one could arrive at a solution where the 
decay-width scaling factor is less than unity and consistent with the 
Kramers' scaling factor in Eq.~(\ref{eq:Kramers}).

\begin{figure}[tbp]
\includegraphics*[width=\linewidth]{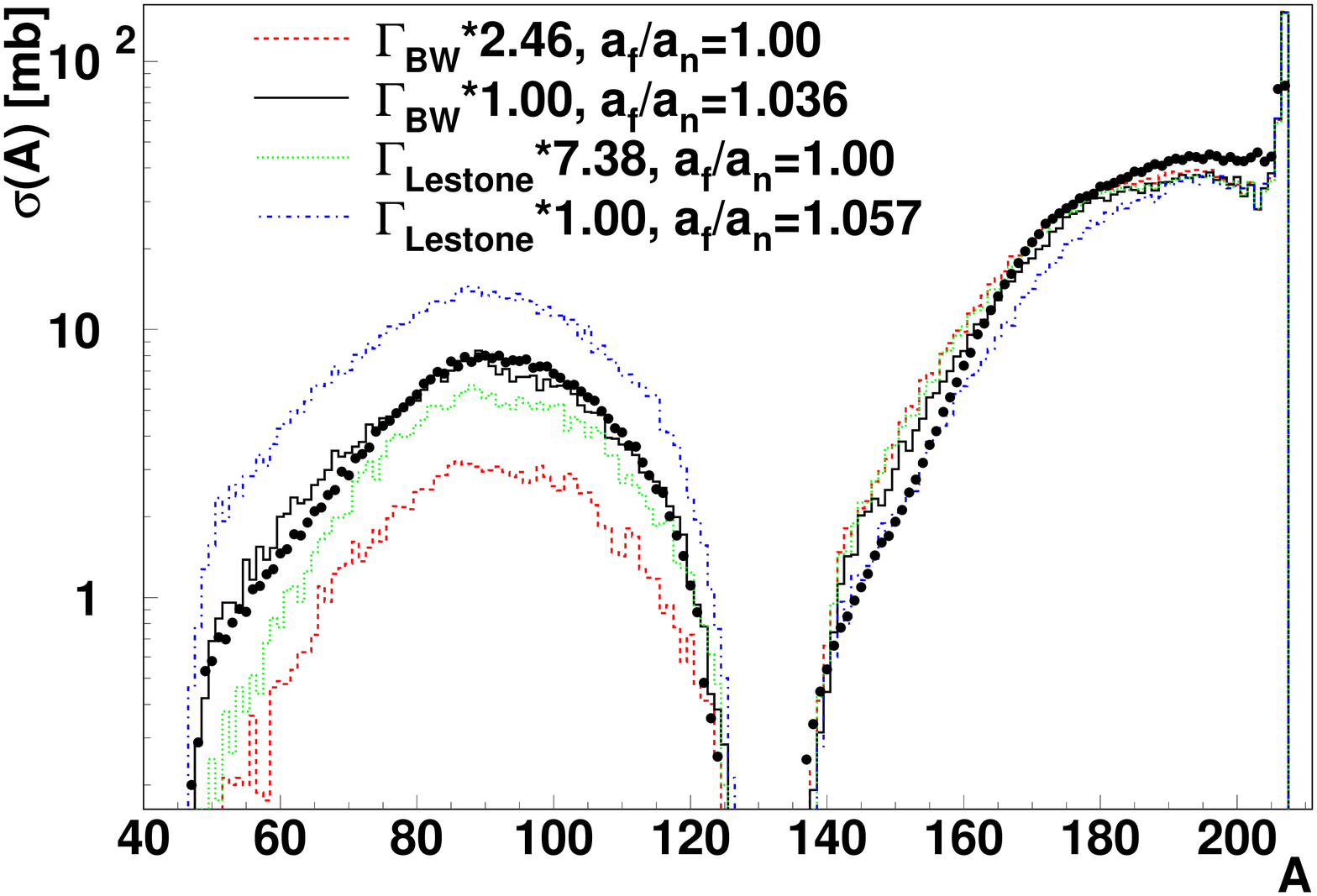}
\caption{(Color online) Comparison of experimental and calculated residual mass distributions for the 1-GeV
  \textit{p}+$^{208}$Pb reaction. Predictions of the \incl-\geminipp\ code are shown for different adjustments of the
  fission width.  Experimental data from Ref.~\protect\cite{enqvist-lead}.}
\label{fig:fitpPb}
\end{figure}

As it is impossible to distinguish these different ways of modifying the fission probability 
from the fusion data alone, we now consider the constraint of adding the spallation data to the analysis.
In Fig.~\ref{fig:fitpPb}, we show the equivalent calculations for the mass distributions of the products 
of the 1-GeV
\textit{p}+$^{208}$Pb spallation reaction. Of all these possibilities, the 
$\Gamma_\text{BW}\times 1.00, a_\text{f}/a_\text{n}=1.036$ calculation reproduces the yield of the fission peak best.
This highlights the significant reduction in fitting-parameter 
ambiguity that can be obtained by
simultaneously fitting heavy-ion and spallation data. We also note
that the $\Gamma_\text{BW}\times 2.46$, $a_\text{f}/a_\text{n}=1.00$ calculation and the reduced fission
barrier calculation with $\Gamma_\text{BW}\times1.00$, $a_\text{f}/a_\text{n}=1.00$ (not shown) were
again identical. Thus, while many of the ambiguities in the
fitting-parameters have been removed, the ambiguity between the
effect of the magnitude of the fission barrier and the Kramers
scaling factor was not not lifted with simultaneous fitting.
%  We also note that the 
% $\Gamma_\text{BW}\times 2.46, a_\text{f}/a_\text{n}=1.00$ calculation and the reduced fission barrier calculation with
% $\Gamma_\text{BW}\times 1.00$, $a_\text{f}/a_\text{n}=1.00$ (not shown) were again almost identical. Thus the ambiguity between 
% the effect of magnitude of the fission barrier and Kramers scaling factor
% was not lifted with the inclusion of spallation. However the other 
% ambiguities associated with our fitting parameters were removed. 
We will continue using the  
$\Gamma_\text{BW}\times 1.00, a_\text{f}/a_\text{n}=1.036$ calculation as our  
best fit to both sets of experimental data, but it should be noted that with 
reduced fission barriers, an equivalent solution with a 
Kramers scaling factor ($<1$) can also be obtained. We are just not able to constrain the magnitude of the Kramers factor from all these data.

\begin{figure}[tbp]
\includegraphics*[scale=0.4]{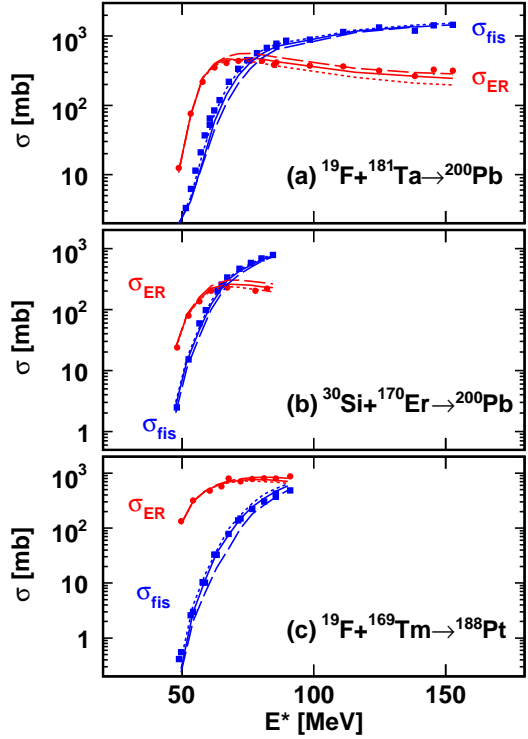}
\caption{(Color online) Comparison of experimental and calculated fission and evaporation-residue excitation functions for
  the indicated reactions. Solid lines: Bohr-Wheeler fission width, $a_\text{f}/a_\text{n}=1.036$, no fission delay. Dashed
  lines: Lestone fission width, $a_\text{f}/a_\text{n}=1.057$, 1-zs fission delay. Dotted lines: Bohr-Wheeler fission
  width, energy-dependent effective $a_\text{f}/a_\text{n}$ ratio with $r$=1.0747.}
\label{fig:fisPb}
\end{figure}

\begin{figure}[tbp]
\includegraphics*[scale=0.4]{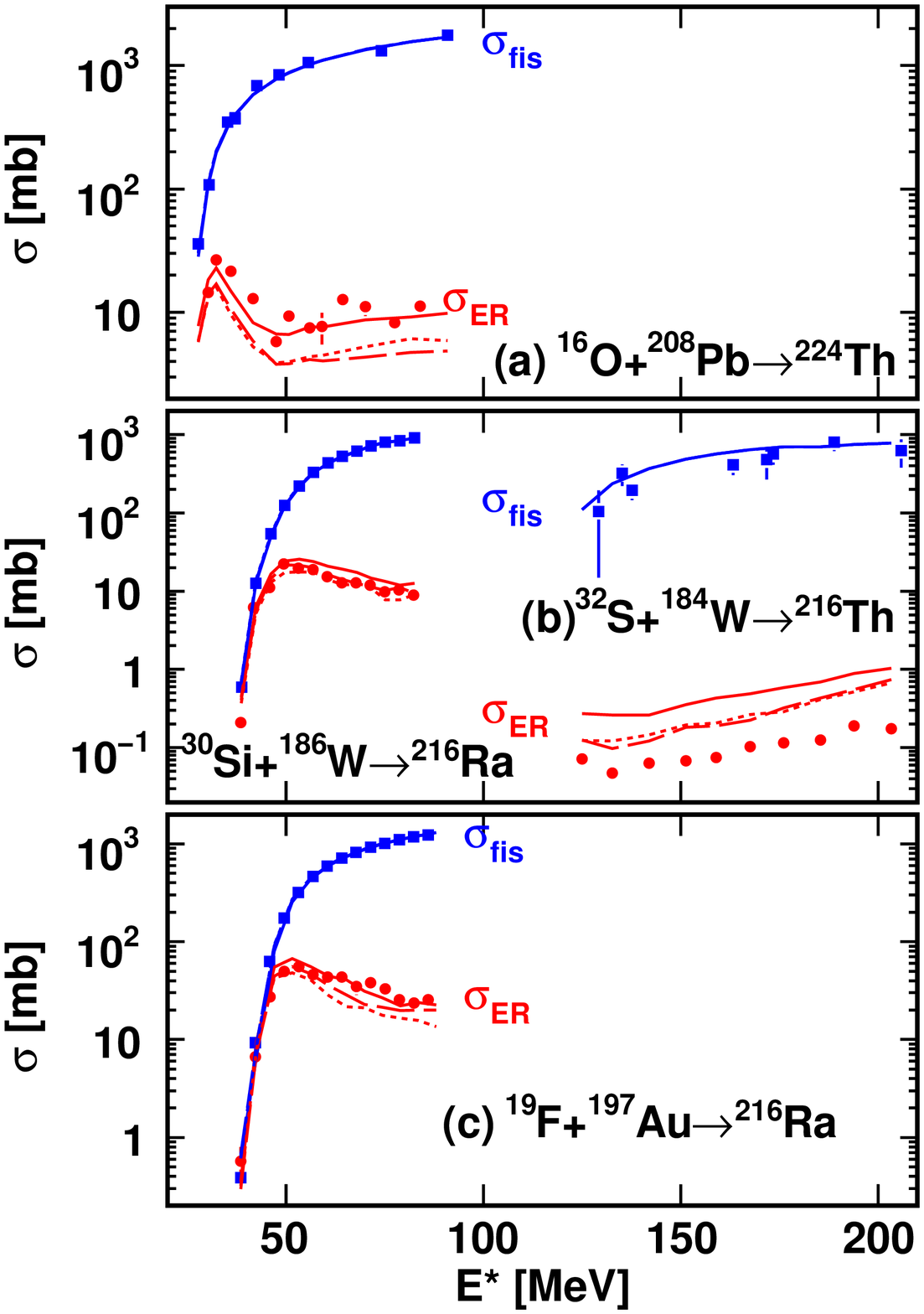}
\caption{(Color online) Same as Fig.~\ref{fig:fisPb}.}
\label{fig:fisTh}
\end{figure}

\begin{figure}[tbp]
\includegraphics*[scale=0.4]{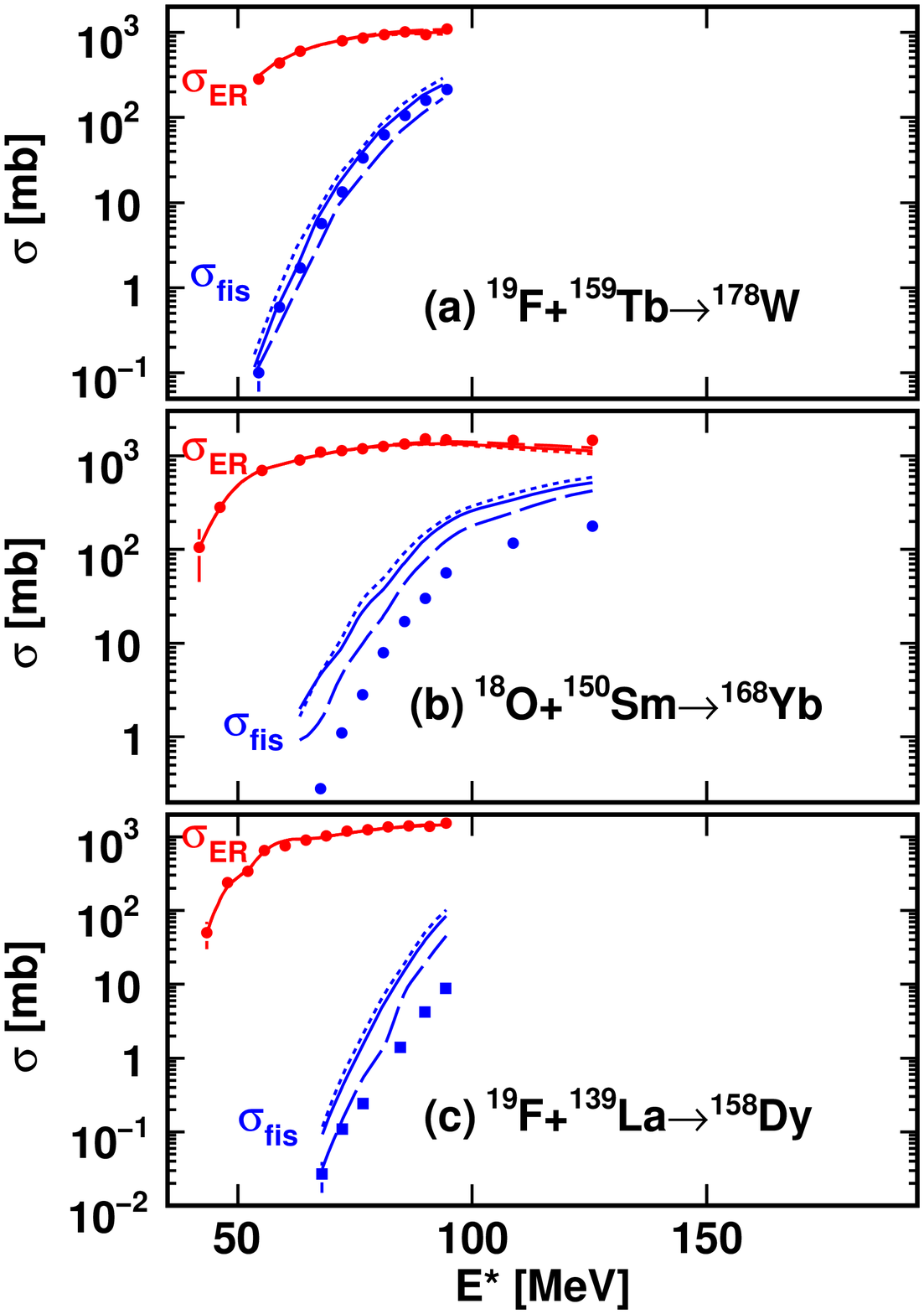}
\caption{(Color online) Same as Fig.~\ref{fig:fisPb}.}
\label{fig:fisYb}
\end{figure}

\begin{figure}[tbp]
\includegraphics*[scale=0.4]{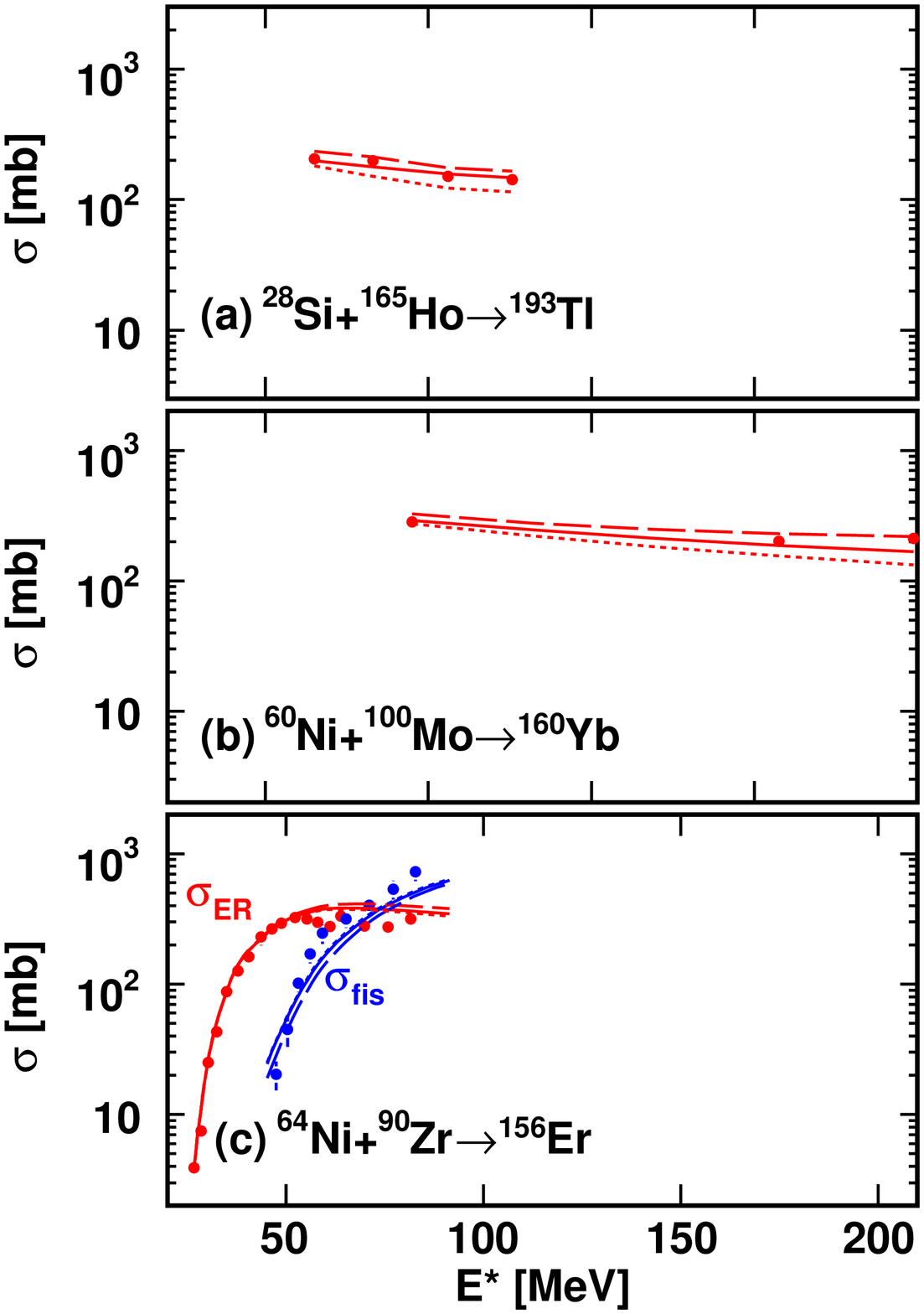}
\caption{(Color online) Same as Fig.~\ref{fig:fisPb}.}
\label{fig:fisTl}
\end{figure}

\begin{figure}
  \includegraphics[width=\linewidth]{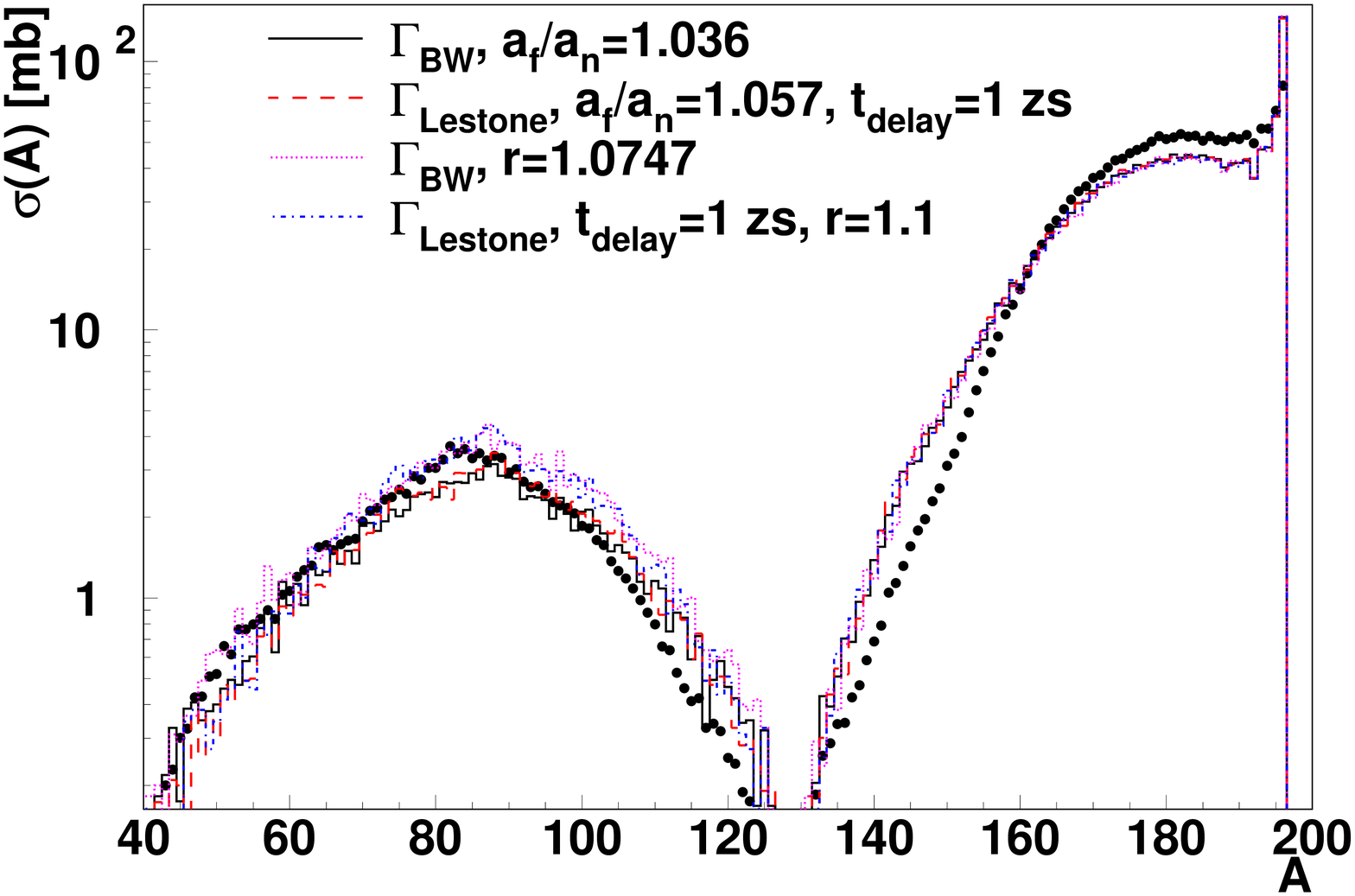}
  \caption{(Color online) Residue-mass distribution for the \textit{p}+$^{197}$Au reaction at 1~GeV. Predictions of the
    \incl-\geminipp\ code are shown for different adjustments of the fission width. %  Solid black line: Bohr-Wheeler fission
    % width, $a_\text{f}/a_\text{n}=1.036$, no fission delay. Dashed red line: Lestone fission width,
    % $a_\text{f}/a_\text{n}=1.057$, 1-zs fission delay. Dotted purple line: Bohr-Wheeler fission width, energy-dependent
    % effective $a_\text{f}/a_\text{n}$ ratio [Eq.~(\ref{eq:fita_r})]. Dot-dashed blue line: Lestone fission width, 1-zs
    % fission delay, energy-dependent effective $a_\text{f}/a_\text{n}$ ratio [Eq.~(\ref{eq:fita_r})].
    Experimental data
    from Ref.~\protect\cite{benlliure-gold}.}
  \label{fg:fission_au}
\end{figure}

\begin{figure}
  \includegraphics[width=\linewidth]{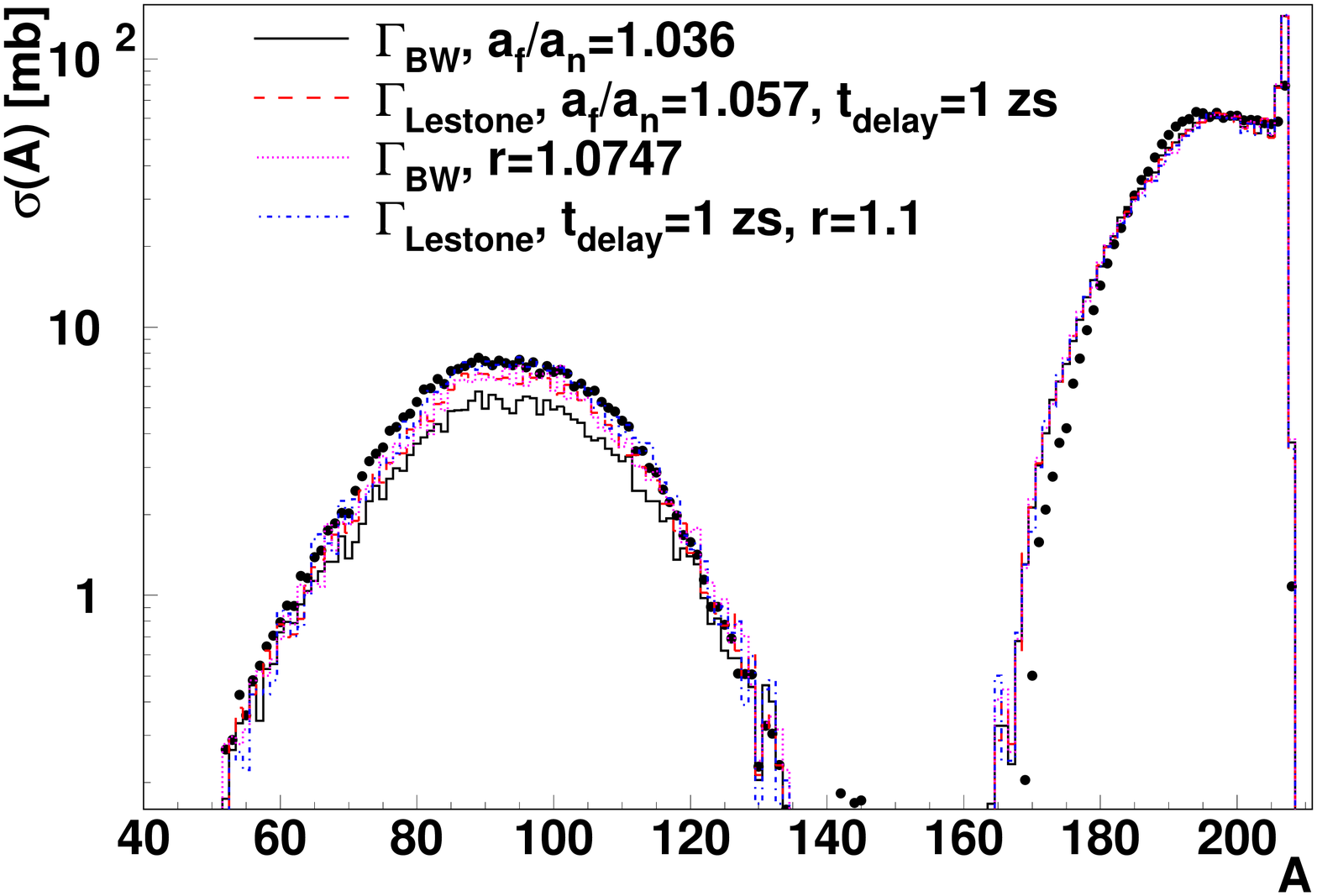}
  \caption{Same as Fig.~\ref{fg:fission_au} for \textit{p}+$^{208}$Pb at 500~MeV. Experimental data from
    Refs.~\protect\cite{audouin-lead,fernandez-lead}. The experimental fission cross sections have 
   been multiplied by a factor of 146/232
    (see text for details).}
  \label{fg:fission_pb500}
\end{figure}

\begin{figure}
  \includegraphics[width=\linewidth]{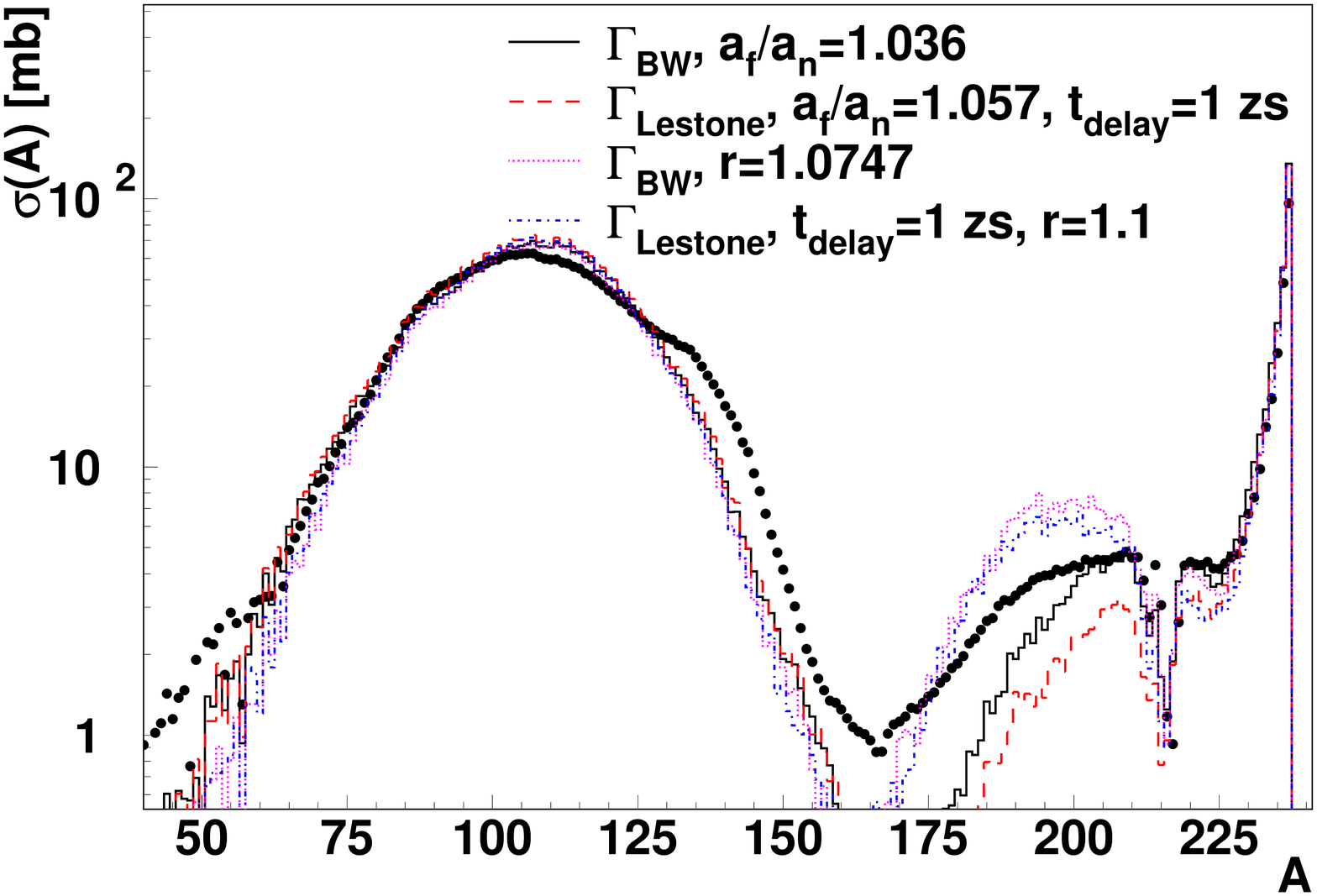}
  \caption{(Color online) Same as Fig.~\ref{fg:fission_au} for \textit{p}+$^{238}$U at 1~GeV. Experimental data from
    Refs.~\protect\cite{bernas-u_fission,taieb-u,ricciardi-uranium}.}
  \label{fg:fission_u}
\end{figure}

\begin{figure}
  \includegraphics[width=\linewidth]{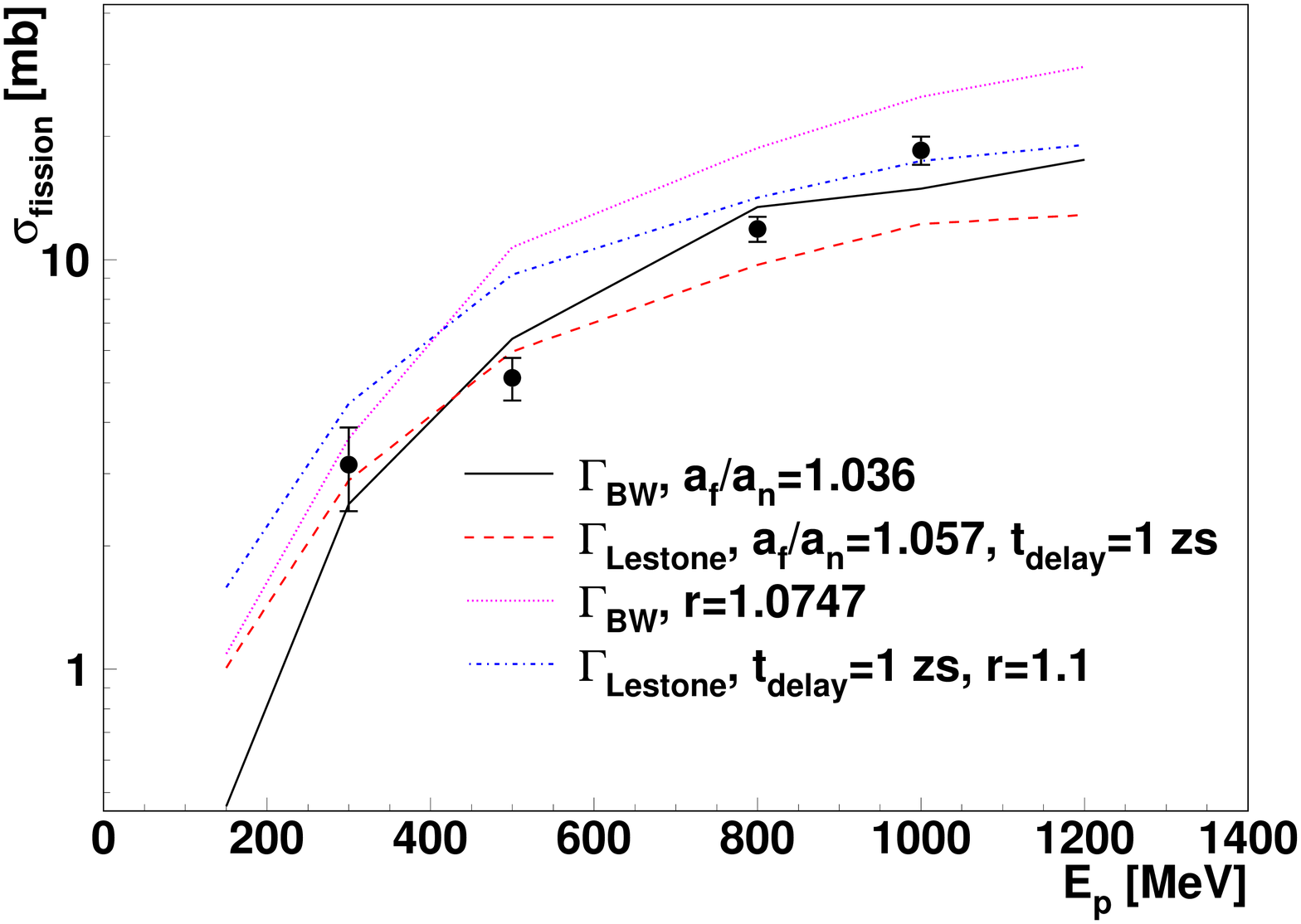}
  \caption{(Color online) Excitation curve for the fission cross section in \textit{p}+$^{181}$Ta. Predictions of the
    \incl-\geminipp\ code are shown for different adjustments of the fission width. % Solid black line: Bohr-Wheeler fission
    % width, $a_\text{f}/a_\text{n}=1.036$, no fission delay. Dashed red line: Lestone fission width,
    % $a_\text{f}/a_\text{n}=1.057$, 1-zs fission delay. Dotted blue line: energy-dependent effective
    % $a_\text{f}/a_\text{n}$ ratio [Eq.~\eqref{eq:fita_r}].
    Experimental data from
    Ref.~\protect\cite{benlliure-ta}.}
  \label{fg:Ta_excit}
\end{figure}

Comparison of \geminipp\ predictions to the experimental fission and evaporation-residue excitations functions listed for
the heavy-ion-induced fusion reactions in Table~\ref{Tbl:fus} are shown in Figs.~\ref{fig:fisPb} to \ref{fig:fisTl}. 
The solid curves show the predictions with $a_\text{f}/a_\text{n}$=1.036, 
Sierk fission barriers, and no scaling of the Bohr-Wheeler decay width.  For
spallation, Figs.~\ref{fg:fission_au} to \ref{fg:fission_u} show the comparison between measured and calculated residue
mass distributions. Finally, Fig.~\ref{fg:Ta_excit} shows the comparison between measured \cite{benlliure-ta} and
calculated excitation curves for the fission cross section in proton collisions with $^{181}$Ta, a low-fissility target.

%\begin{figure}
%  \centering
%  \includegraphics[width=0.8\linewidth]{fission_pb1000}
%  \caption{Same as Fig.~\ref{fg:fission_au} for \textit{p}+$^{208}$Pb at 1000~MeV.}
%  \label{fg:fission_pb1000}
%\end{figure}

The central result is that it is possible to reproduce the total fission cross section for all the studied spallation
reactions by adjusting only one free parameter, namely the $a_\text{f}/a_\text{n}$ ratio, which was set equal to $1.036$ in
our calculations, whilst the global scaling of the fission width and of the fission barrier were kept equal to $1$; no
Lestone correction was introduced. A global scaling of the fission width is 
roughly equivalent to a reduction of the barrier
height, but in both cases, these adjustment alone do not fit the data. 
The adjustment of the $a_\text{f}/a_\text{n}$ ratio, on the other
hand, is characterized by a different excitation-energy dependence, which is better suited for the description of fission
from spallation remnants. The Lestone correction, which suppresses the fission width at high spin, does not have a large
effect on spallation data, some 80\% at most. This is due to the small angular momenta generated in the intranuclear
cascade.

For the heavy-ion-induced fusion data,
the \geminipp\ predictions are also generally quite good, however, there are a couple of reactions where significant 
deviations are found. Firstly, for the $^{18}$O+$^{150}$Sm$~\rightarrow^{168}$Yb and 
$^{19}$F+$^{139}$La$\rightarrow^{158}$Dy reactions shown in Figs.~\ref{fig:fisYb}(b) and \ref{fig:fisYb}(c), 
the fission cross section is 
overestimate by almost an order of magnitude. In comparison, the Ni-induced reactions in Figs.~\ref{fig:fisTl}(b) and
\ref{fig:fisTl}(c)
making similar mass compound nuclei are reproduced much better. It is  difficult to understand these
$^{18}$O- and $^{19}$F-induced reactions and previous attempts also failed to reproduce the data \cite{Charity86}. 
For instance at the highest bombarding energies, the  excitation energies  
probed in these reactions overlap those in the Ni-induced reactions. Similarly the predicted angular-momentum region 
over which the
fission yield is determined in the O- and F-induced reactions is similar to that in which the residue yield 
is determined in the Ni-induced reactions. Unless there are significant non-fusion processes, such as pre-equilibrium or
incomplete fusion occurring, these data  suggest an entrance-channel dependence of the fission decay 
probability which would violate the compound-nucleus hypothesis. However we see no evidence of such an effect 
for the O- and F-induced induced reactions with heavier targets. Clearly our understanding of fission for 
$A<{}$170 is lacking and more studies are needed. 

The other case where the \geminipp\ predictions fail is for the $^{32}$S+$^{184}$W$\rightarrow^{216}$Th reaction in 
Fig.~\ref{fig:fisTh}(a). Here the evaporation-residue cross section is exceedingly small ($\sim{}$0.1~mb) and is 
overpredicted by almost an order of magnitude. However, the calculations gets the excitation-energy dependence of the 
cross sections correct which previous calculations could not do without invoking an excitation-energy dependence 
of the dissipation strength \cite{Back99}. In our calculations the predicted excitation-energy dependence of the 
residue cross section is a consequence
of the assumed excitation-energy dependence of the level-density parameter \cite{Charity10}.
Low probability events in the statistical model are 
generally quite sensitive to the statistical-model parameters. In this case, it was demonstrated that the residue yield
is very sensitive the absolute value of the level-density parameter and its excitation-energy dependence 
\cite{Charity10}. For example, the residue yield is increased  by 2-3 orders of magnitude when the level-density
parameter is changed from $a=A/7.3$ to $A/11$~MeV$^{-1}$.  Further refinement of the value of this parameter at the 
larger excitation energies probed in this more symmetric fusion reaction may be needed in the future.

Alternatively,
there is evidence that for this mass region, quasi-fission completes with fusion reactions even at the lower $\ell$ waves 
associated with evaporation residue production \cite{Berriman01,Hinde02}.  In the case of $^{216}$Ra compound nucleus in 
Figs.~\ref{fig:fisTh}(b) and \ref{fig:fisTh}(c), Berriman \etal\ have indicated that 
both the $^{19}$F+$^{197}$Au 
and $^{30}$Si+$^{184}$W reactions
have reduced evaporation residue cross sections due to quasi-fission competition \cite{Berriman01}. Therefore this higher mass region
for the heavy-ion reactions is subject to more uncertainty in constraining the statistical-model parameters. 

\subsection{Fission-Fragment Mass Distributions}
\label{sec:fmass}

Previous treatments of the  fission-fragment mass distribution have assumed
thermal models where the mass division is determined at either 
the saddle-point (Moretto's formalism) or at the scission configuration \cite{Wilkins76}. In reality, a complete description probably 
requires one to follow the trajectory from saddle to scission including fluctuations, for example by 
Langevin simulations \cite{Karpov01,Ryabov09}. However for large-scale simulations, this is too time consuming, so a simpler 
and faster procedure for determining mass division is required.

\begin{figure}[tbp]
\includegraphics*[scale=0.4]{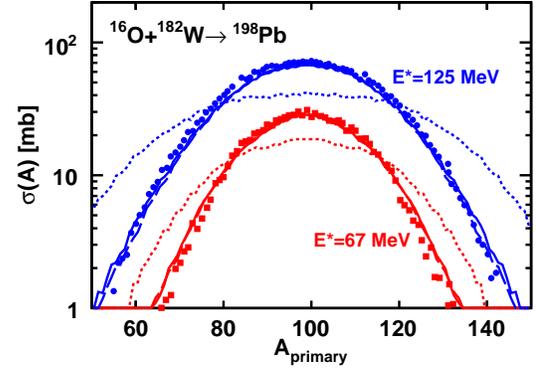}
\caption{(Color online) Comparison of experimental (data points) and simulated (curves) distributions of
the primary fission-fragment masses for the $^{198}$Pb compound nucleus formed in 
the $^{16}$O+$^{182}$W reaction at the two indicated excitation energies. The total fission cross section has been
calculated using the Bohr-Wheeler formalism, $a_\text{f}/a_\text{n}=1.036$, without any fission delay.
The dotted curves was obtained using the Moretto formalism with Sierk's conditional barriers to 
define the mass distributions. 
The solid and dashed curves were obtained using the Rusanov systematics 
using the saddle and scission-point temperatures, respectively.}
\label{fig:fmass198}
\end{figure}

\begin{figure}[tbp]
\includegraphics*[scale=0.4]{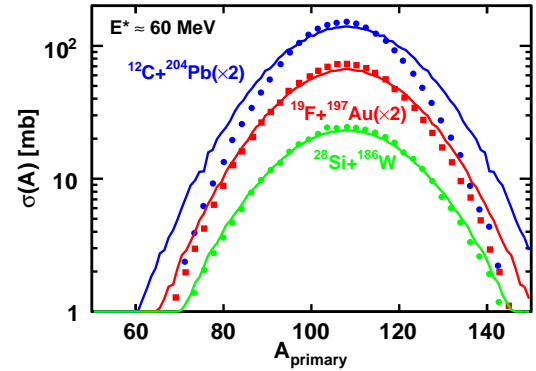}
\caption{(Color online) Comparison of experimental (data points) and simulated (curves) distributions of
the primary fission-fragment masses for the $^{216}$Ra compound nucleus at $E^*\sim60$~MeV formed in 
the $^{12}$C+$^{204}$Pb, $^{19}$F+$^{197}$Au and $^{30}$Si+$^{186}$W 
reactions. The total fission cross section has been calculated
using the Bohr-Wheeler formalism, $a_\text{f}/a_\text{n}=1.036$, without any fission delay.
To aid in viewing, the data and curves have been scaled by the 
indicated amounts.}
\label{fig:fmass216}
\end{figure}

Experimental mass distributions for  heavy-ion-induced fusion-fission reactions are shown in
Figs.~\ref{fig:fmass198} and \ref{fig:fmass216}. In these figures, the experimental fission-fragment 
masses was not directly measured, but rather the ratio of primary masses 
(before post-scission particle evaporation) was inferred from either the ratio of the measured 
fission-fragment velocities or kinetic energies.
The absolute primary mass was assumed to be equal to the compound-nucleus mass, 
which of course ignores the pre-scission evaporation of light particles. 
However, the distributions
simulated by \geminipp\ (the curves in the figures) were analyzed in the same manner as the 
experimental data and thus contain the same deficiencies. 

The data in Fig.~\ref{fig:fmass198} is for the 
$^{16}$O+$^{182}$W$\rightarrow^{198}$Pb reactions. The relative mass distribution were obtained 
from Ref.~\cite{Plasil66} and absolute normalization was achieved using the fission cross sections 
measured in Ref.~\cite{Sikkeland64}.  The data in 
Fig.~\ref{fig:fmass216} are for the $^{216}$Ra compound nucleus at $E^*\sim60$~MeV with three 
entrance channels: $^{12}$C+$^{204}$Pb, $^{19}$F+$^{197}$Au, and $^{28}$Si+$^{186}$W. 
The fission excitation function for the 
latter two are shown in Figs.~\ref{fig:fisTh}(b) and \ref{fig:fisTh}(c).

In Fig.~\ref{fig:fmass198}, the dotted curves show
the mass distribution determined from Moretto's formalism using interpolated values of Sierk's 
finite-range calculations for the conditional barriers \cite{charity-gemini++}. The total fission width (the total width for all
mass divisions associated with the peak in the mass distribution) was normalized to the Bohr-Wheeler 
value for these calculations. Therefore in this figure, only the shape of the mass distribution is 
determined from the Moretto 
formalism. Clearly these distributions are much wider than the experimental quantities.
This is quite typical of other cases where a peak exist in the mass distribution at symmetry.
See for example the study of $^{151}$Eu compound nuclei 
in Ref.~\cite{charity-Eu} and the light, but high spin $^{110}$Sn compound nuclei
studied in Ref.~\cite{sobotka-Sn}. For less fissile nuclei where the mass distribution has a 
minimum at symmetric division, the Moretto formalism (with Sierk barriers) give a much better description of the 
experimental data. See for example the studies of  $^{111}$In \cite{sierk-asyBar}, $^{102}$Rh and $^{105}$Ag \cite{charity-gemini} and $^{75}$Br \cite{han-Br} compound nuclei.

The cause of this inadequacy for the heavier systems could either be an incorrect asymmetry dependence of Sierk's 
conditional barriers or a failure of Moretto's formalism. The latter predicts the  
asymmetry distribution at the ridge line of conditional saddle points
and assumes that the mass asymmetry is unchanged during the descent from saddle to scission. 

As an alternative to using Moretto's formalism, we have used the systematics of fission-fragment
mass distributions complied by Rusanov \etal\ \cite{rusanov-mass}. The mass distribution is assumed to be Gaussian and its
variance
is parameterized as
\[
\sigma_A^2 = \frac{A_{CN}^2 T}{ 16 \frac{\ud^2V}{\ud\eta^2}(Z^2/A,J)}
\]
where $\frac{\ud^2V}{\ud\eta^2}$ is the second derivative of the potential energy surface with respect to the 
mass-asymmetry deformation parameter($\eta=2\frac{A_1-A_2}{A_1+A_2}$ where $A_1$ and $A_2$ is the mass division). 
This quantity is parameterized as a function of the fissility $Z^2/A$ and spin $J$. 
The quantity $T$ is the nuclear temperature where 
\[
\frac{1}{T}=\frac{\ud \ln\rho}{\ud U}.
\]

Rusanov \etal\ considered three parameterizations of  $\frac{\ud^2V}{\ud\eta^2}$, with three different 
temperatures. 	Either 1) the temperature of the fission nucleus at the saddle-point is used, 
but no pre-saddle light-particle evaporations are allowed, 
2) as above, but pre-saddle evaporations are allowed, or 
3) the temperature at the scission point is used.  
The first of these is not realistic and was not considered and the second is basically consistent with the ideas 
of the Moretto formalism. The latter two can be called saddle-point 
and scission-point models where the mass distributions are both determined thermally. In these two cases, the
quantity  $\frac{\ud^2V}{\ud\eta^2}$ should be identified with the asymmetry dependence of the potential-energy surface at the saddle and scission points, 
respectively.

In \geminipp, once fission is decided for an event, evaporation during 
the saddle-to-scission transition is allowed. This is important for the scission model,
 as we need to determine the temperature at the scission point.
The saddle-to-scission evaporation is treated in a simplified manner using 
spherical level densities and transmission coefficients in
the Weisskopf-Ewing evaporation formalism 
with the deformation-plus-rotational energy removed from the total excitation energy. The 
deformation-plus-rotational energy of 
the scission configuration is  determined as the sum of fission-fragment kinetic energy from Viola's 
systematics \cite{Viola85} and the fission Q-value. 
Evaporation during the saddle-to-scission transition occurs for a 
period proportional to the difference in energy between the saddle and scission 
points, i.e.
\[
t_\text{ss}= k_\text{ss} (E_\text{saddle}-E_\text{scission})
\]
consistent with large viscosity.  The parameter $k_\text{ss}$ is related to the magnitude of this viscosity was fixed to $k_\text{ss}$=1~zs/MeV by fitting pre-scission neutron multiplicities from 
Ref.~\cite{hilscher-prescissionN}.

\begin{figure}[tbp]
\includegraphics*[width=\linewidth]{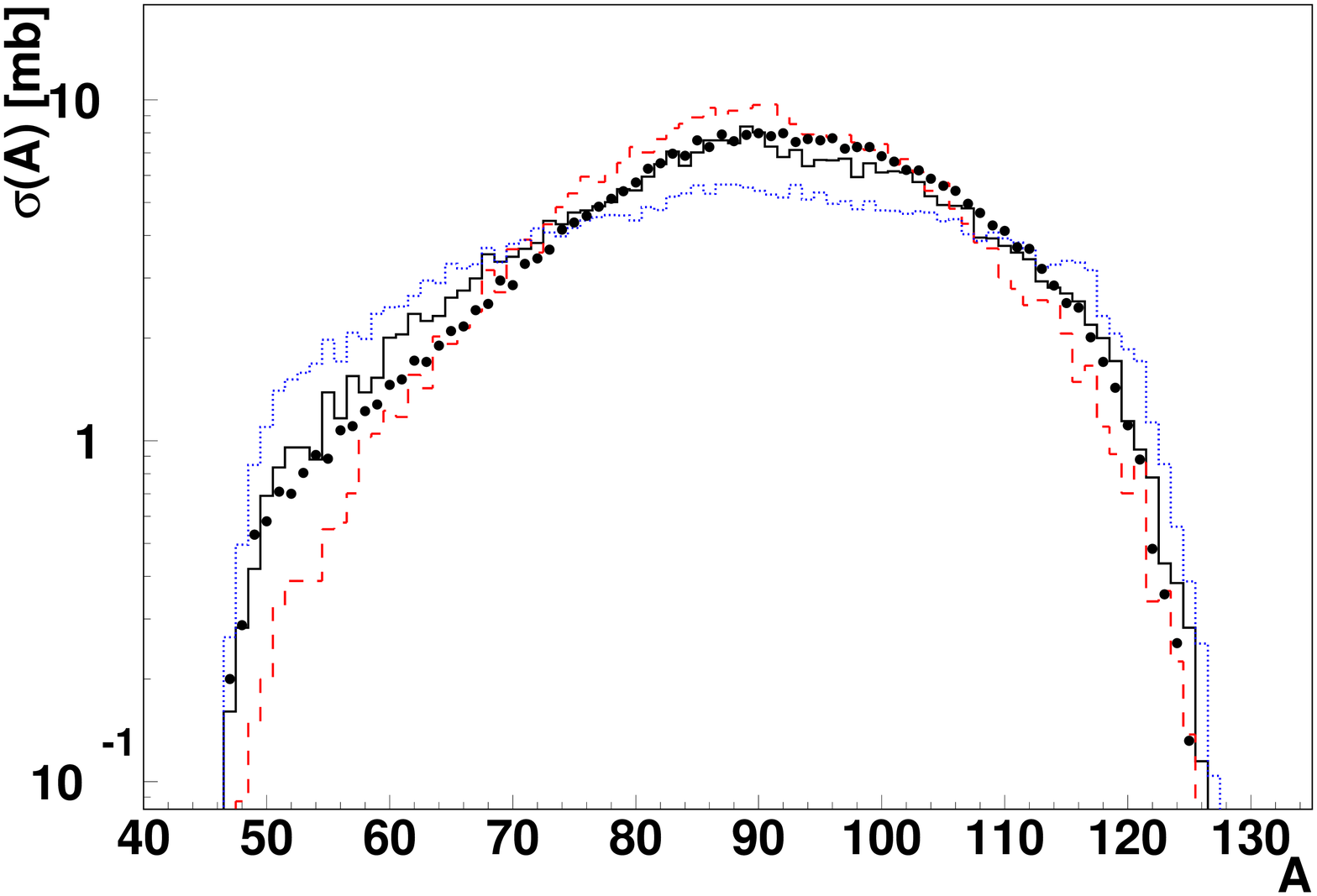}
\caption{(Color online)  Comparison of experimental (data points) and simulated (curves) distributions of
the fission-fragment masses for the 1-GeV \textit{p}+Pb spallation reaction.
The blue dotted curve was obtained using the Moretto formalism with Sierk's conditional barriers to 
define the mass distributions. 
The black solid and red dashed curves were obtained using the Rusanov systematics 
using the saddle and scission-point temperatures, respectively.}
\label{fig:fmassPb}
\end{figure}

The solid and dashed curves in Fig.~\ref{fig:fmass198} show the predictions with the Rusanov saddle 
and scission-point systematics, respectively. These predictions are almost identical, and for the 
lowest excitation energy, the curves completely overlap and cannot be distinguished. 
This is not surprising as both Rusanov systematics are fits to $^3$He induced fission and 
fusion-fission data including the data set of Fig.~\ref{fig:fmass198}. However, at higher 
excitation energies such as those sampled in spallation reactions, the two systematics give 
quite different predictions as the thermal excitation at scission increases much more slowly with 
compound-nucleus excitation than does the saddle-point value \cite{hilscher-prescissionN}.
Figure~\ref{fig:fmassPb} compares, the two systematics for the 1-GeV p+Pb spallation reaction. In this
case, the predicted mass distribution obtained with the scission systematics (dashed curve) is 
too narrow, while the saddle systematics (solid curve) gives good agreement.

The success of Rusanov's saddle-point systematics thus suggests that the
 fission mass division is determined quite close to the saddle-point 
configuration. It addition it indicates that the Moretto formalism is 
still applicable for near symmetric divisions of heavy nuclei. 
However it should not be used with Sierk's conditional barriers in this region.

\begin{figure}[tbp]
\includegraphics*[scale=0.4]{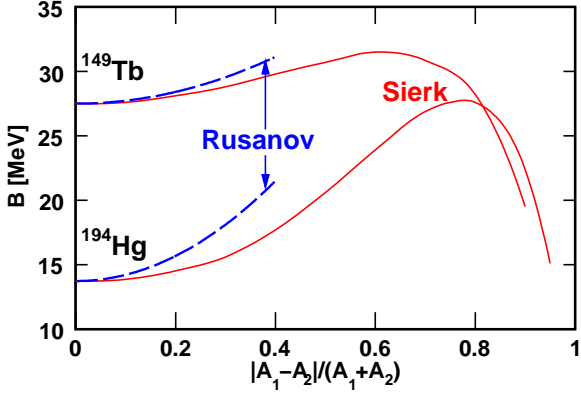}
\caption{(Color online) Comparison of asymmetry dependences of conditional barriers for $^{149}$Tb and
$^{194}$Hg nucleus at $J$=0. The solid curves are the predictions from Sierk's finite-range calculations.
The dashed curves are parabolic functions with curvatures taken from the 
Rusanov systematics and with the symmetric fission barriers taken from the finite-range calculations.
The mass-asymmetry coordinate is defined in terms of $A_{1}$ and $A_{2}$, the two masses following binary division. } 
\label{fig:sierk}
\end{figure}

The differences between the mass-asymmetry dependence of Sierk's
conditional barriers and the Rusanov systematics are shown directly 
in Fig.~\ref{fig:sierk} for $^{149}$Tb and $^{194}$Hg compound nuclei at $J$=0.
The dashed curves are parabolic functions with curvatures from the Rusanov systematics 
and with the symmetric fission barriers from Sierk's calculations. The Rusanov results  
have larger curvatures at symmetry than Sierk's predictions and thus give narrower 
fission-fragment mass distributions. The differences between Sierk's predictions and 
the Rusanov systematics is much larger for the heavier $^{194}$Hg nucleus. 
For even heavier nuclei, the asymmetry coordinate in the finite-range calculations
becomes undefined as the saddle-point configuration has no well defined neck \cite{Thomas85}.
If this is the case, then the Moretto formalism is no longer applicable for these systems 
and the mass asymmetry is determined during the descent from saddle to scission. 
In such cases the interpretation of the Rusanov systematics in terms of a 
Moretto-type may be suspect. We note that the 
$Z^2/A$ dependence of $\ud^2V/\ud\eta^2$ in Rusanov systematics has an
abrupt slope change at $Z^2/A$=24 possibly related to this effect. 
However, even in the \textit{p}+U spallation reaction we produce $Z^2/A$ ratios 
that are below this value.  
 
The Rusanov saddle systematics was used for the 
other spallation predictions in Figs.~\ref{fg:fission_au} to \ref{fg:fission_u} and gives quite good agreement.
 However, 
for the \textit{p}+$^{238}$U reaction in Fig.~\ref{fg:fission_u}, the simulation fails to reproduce the small shoulder 
in the fission mass distribution for higher mass. The Rusanov systematics only gives the width of the distribution and will not 
predict finer structures linked to shell effects, such as this.

In Fig.~\ref{fig:fmass216},  the simulated mass distributions (from the saddle systematics) for the $^{216}$Ra compound 
nuclei reproduce the data 
reasonably well with the exception of the $^{12}$C+$^{204}$Pb data where 
experimental distribution is somewhat narrower. 
Berriman \etal\ \cite{Berriman01} suggest that $^{12}$C+$^{204}$Pb data is all 
fusion-fission while the $^{19}$F+$^{197}$Au and $^{30}$Si+$^{184}$W data both contain quasi-fission contributions making the 
mass distributions wider. This would imply that for the more massive compound nucleus, the Rusanov systematics 
overestimate the width of the statistical fission mass distributions as many of the heavy-ion data used in these 
systematics have contributions from quasi-fission. However, the Rusanov systematics also
contains the lower-spin $^{3}$He-induced fission data in this mass region and here quasi-fission is expected to be absent. 
Thus spallation mass distributions which sample lower spins are not expected to suffer from this problem.

\subsection{Fission Delays}
     
\begin{figure}[tbp]
\includegraphics*[width=\linewidth]{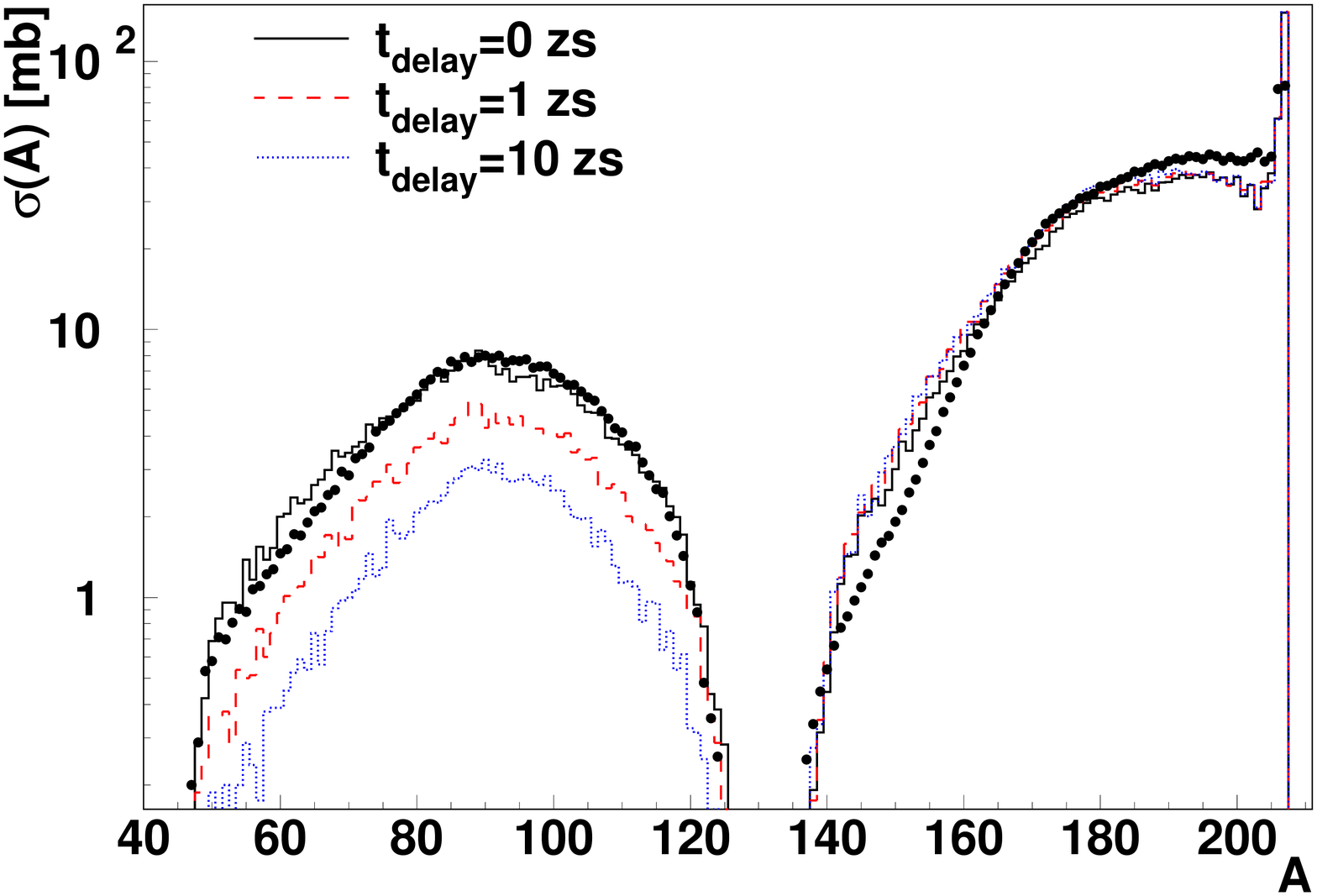}
\caption{(Color online) Same as Fig.~\ref{fig:fitpPb} for the Bohr-Wheeler fission width, $a_\text{f}/a_\text{n}=1.036$ and
    three different values of the fission delay.}
\label{fig:tdelay}
\end{figure}
\begin{figure}[tbp]
  \includegraphics[width=\linewidth]{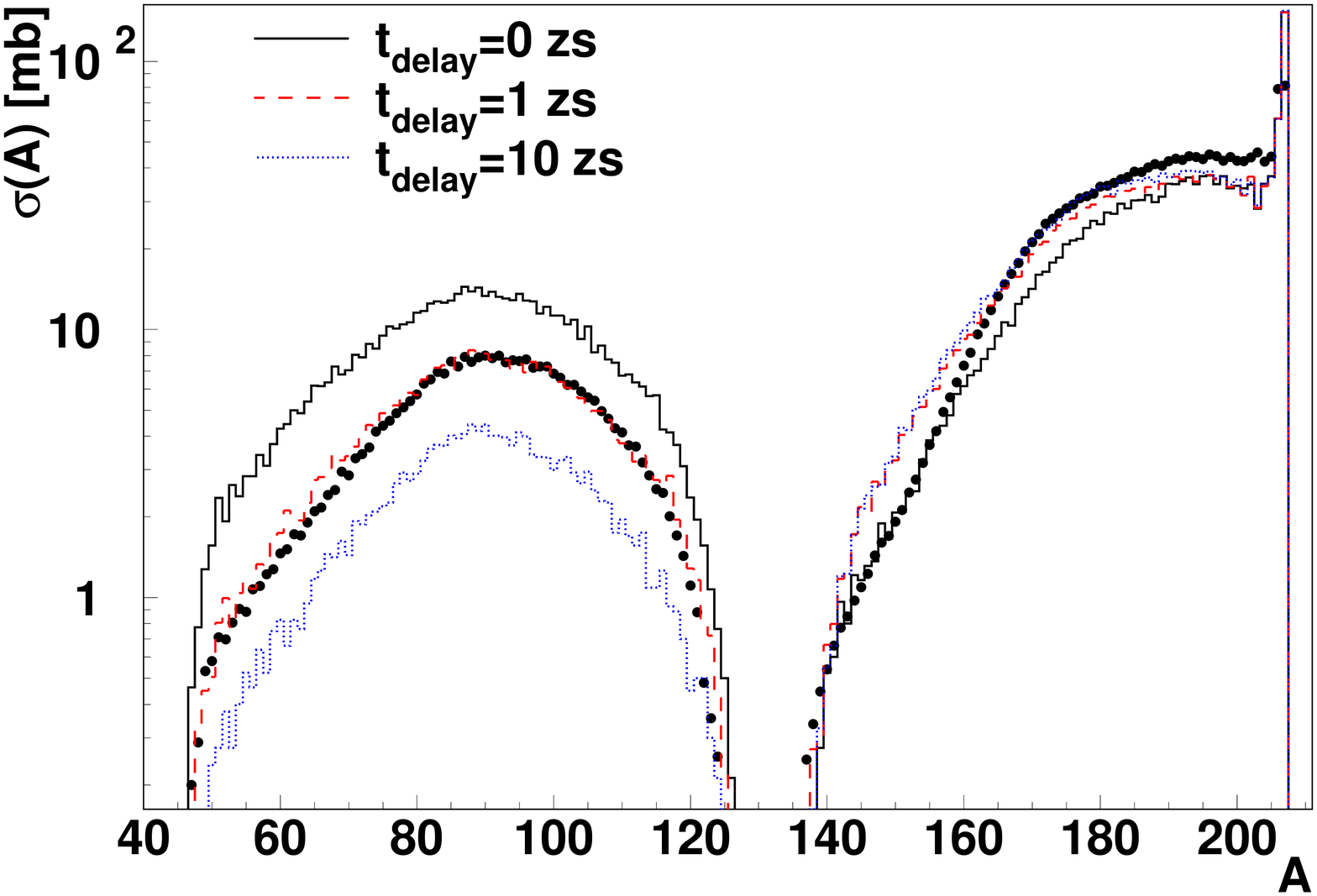}
  \caption{(Color online) Same as Fig.~\ref{fig:fitpPb} for the Lestone fission width, $a_\text{f}/a_\text{n}=1.057$ and
    three different values of the fission delay.}
  \label{fg:pb1000_delay}
\end{figure}

Apart from the lighter compound nuclei, we have demonstrated that a unified description of fission widths in fusion and
spallation reactions can be obtained. The solution is unique, apart from an ambiguity between the height of the fission
barrier and the Kramers scaling factor. However we will now show that another ambiguity arises when fission transients are considered.  To show the sensitivity of predictions to transients, we have incorporated a simple implementation
of these in \geminipp; the fission width is set to zero for a time $t_\text{delay}$, after which it assumes its asymptotic
value. During this fission-delay period, the compound nucleus can decay by light-particle evaporation and
intermediate-mass-fragment emission. The fission delay is expected on theoretical grounds to be logarithmically dependent
on nuclear temperature \cite{Grange83}, but this weak dependence (and any mass dependence) has been neglected in
a first approximation. Figure~\ref{fig:tdelay} shows the dependence of the predicted mass distributions for the 1-GeV
\textit{p}+$^{208}$Pb reaction with $t_\text{delay}$=0, 1, and 10~zs. Even a small 1~zs delay has a large effect on the yield in
the fission peak. Therefore, the spallation reactions should be quite sensitive to the fission transients.  Tishchenko
\etal\ also expected large reductions is the fission probability in 2.5-GeV \textit{p}+$^{197}$Au, $^{209}$Bi, $^{238}$U reactions due to
fission transients; however, they were also able to reproduce the fission yield within the standard statistical-model
framework \cite{Tishchenko05}.

\begin{figure}[tbp]
\includegraphics*[scale=0.4]{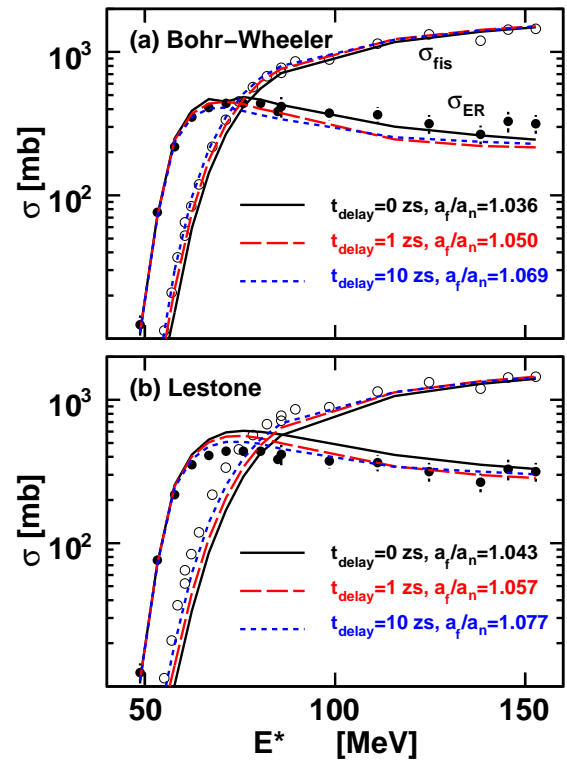}
\caption{(Color online) Comparison of \geminipp\ predictions using (a) the Bohr-Wheeler and (b) the Lestone 
fission formalism to the experimental evaporation-residue and fission excitation functions 
for the $^{19}$F+$^{181}$Ta reaction. The curves are labeled by the $t_\text{delay}$ and $a_\text{f}/a_\text{n}$
values obtained from fitting the fission cross section for the 1-GeV \textit{p}+Pb reaction.}
\label{fig:delay}
\end{figure}

Jing \etal\ \cite{jing-transients} find the effect of increasing the
fission delay can be largely counteracted by increasing the value of the  
$a_\text{f}/a_\text{n}$ parameter. 
Both parameters have little effect on the fission probability 
at low excitation energies. However with increasing excitation energy, 
the fission probability becomes ever more sensitive to both $t_\text{delay}$ and
$a_\text{f}/a_\text{n}$. Even with the large range of excitation energies
explored  in this work, we found it is impossible to break the 
ambiguity between $t_\text{delay}$ and $a_\text{f}/a_\text{n}$.
To illustrate this, Fig.~\ref{fig:delay} compares the $^{200}$Pb 
fusion data to \geminipp\ calculations with fission delay for both the 
Bohr-Wheeler [Fig.~\ref{fig:delay}(a)] and Lestone  [Fig.~\ref{fig:delay}(b)]
formalisms. The values of $t_\text{delay}$ and $a_\text{f}/a_\text{n}$ listed 
in these figures were obtained by reproducing the fission cross section in the 
1-GeV p+Pb spallation reaction. For the Bohr-Wheeler case in 
Fig.~\ref{fig:delay}(a), one see that the calculations with $t_\text{delay}=1$ 
and $10$~zs are almost identical and within  30\% of the experimental values. 
The calculation with $t_\text{delay}=0$~zs fits the data somewhat better, but all
calculations can be deemed acceptable.

For the Lestone formalism in Fig.~\ref{fig:delay}(b), the inclusion of a delay 
with $t_\text{delay}>1$~zs improves the agreement with the data. As 
in Fig.~\ref{fig:delay}(a), the calculations with $t_\text{delay} \geq 1$~zs are 
again almost identical. The Lestone prescription with fission delay also allows 
good agreement with the other data sets we have considered, see the dotted-curves in 
 Figs.~\ref{fig:fisPb} to \ref{fig:fisTl} (fusion) and Figs.~\ref{fg:fission_au} to \ref{fg:fission_u} (spallation)
which were obtained with $a_\text{f}/a_\text{n}=1.057$ and $t_\text{delay}$=1~zs. Calculations with the larger 
$t_\text{delay}$ values produce a similar level of agreement.
It is thus clear that the magnitude of the fission transients cannot be deduced from the fission 
probability alone.

\section{Fission at Very High Excitation Energy}\label{sec:veryhigh}

Fission cross sections in fusion and spallation reactions are dominated by the most densely populated regions of the
compound-nucleus $E^*$-$J$ plane (Fig.~\ref{fig:ejmap}). The successful reproduction of these data thus indicates that the
\geminipp\ model gives an efficient description of fission from compound nuclei with excitation energy up to $\sim300$~MeV
and spin up to $\sim60$~$\hbar$.

It is possible to probe beyond this region if one considers other types of data. Tishchenko \etal\
\cite{tishchenko-fastfission} studied proton-induced spallation reactions at 2.5 GeV on gold, bismuth and uranium
targets. They measured the fission probability in coincidence with neutron, hydrogen and helium multiplicities, which can
be used to reconstruct the excitation energy after the intranuclear cascade. They were able to reproduce the measurements
with an old version of the \code{INCL}-\gemini\ model by tuning the value of $a_\text{f}/a_\text{n}$ on a system-by-system
basis, ranging from $1.000$ for the uranium target to $1.022$ for the gold target. These $a_\text{f}/a_\text{n}$ values are smaller than
those discussed in the present work.

\begin{figure}[tbp]
  \includegraphics[width=\linewidth]{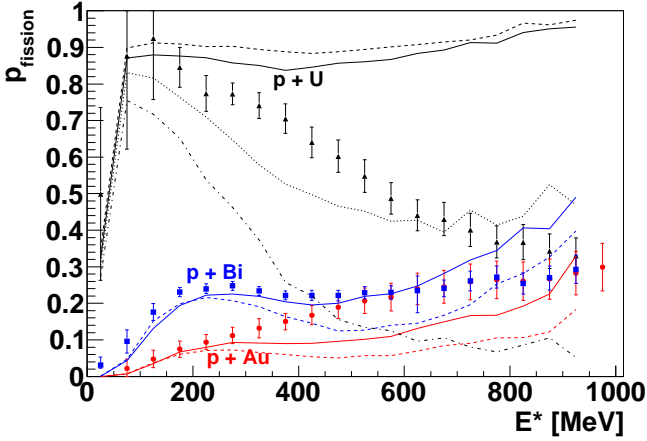}
  \caption{(Color online) Fission probability as a function of the excitation energy of remnants in 2.5-GeV \textit{p}+Au,
    Bi and U reactions. Solid lines: Bohr-Wheeler width, $a_\text{f}/a_\text{n}=1.036$. Dashed lines: Lestone,
    $a_\text{f}/a_\text{n}=1.057$, $t_\text{delay}=1$~zs. Dotted line: Bohr-Wheeler width,
    $a_\text{f}/a_\text{n}=1.02$. Dash-dotted line: Bohr-Wheeler width, $a_\text{f}/a_\text{n}=1.00$.}
  \label{fg:tishchenko_paper1}
\end{figure}

Fig.~\ref{fg:tishchenko_paper1} indeed shows that our candidate parameter sets (Bohr-Wheeler,
$a_\text{f}/a_\text{n}=1.036$; Lestone, $a_\text{f}/a_\text{n}=1.057$, $t_\text{delay}=1$~zs) largely overestimate the
fission probability at high excitation energy deduced by Tishchenko \etal\ for \textit{p}+U. Note that the fission
probability is well reproduced up to a few hundred MeV, which is coherent with the results of the previous section. The
shape of the curve is indeed very sensitive to the value of $a_\text{f}/a_\text{n}$, as Fig.~\ref{fg:tishchenko_paper1}
shows. We can interpret this result an indication of the fact that while a large value of $a_\text{f}/a_\text{n}$ value is
appropriate at low excitation energy, the effective $a_\text{f}/a_\text{n}$ value at high excitation energy should be smaller.

An energy-dependent $a_\text{f}/a_\text{n}$ ratio can naturally appear, among other things, 
as a consequence of the fade out of long-range correlations. To obtain a better reproduction 
of the the Tishchenko data, we have considered a simple refinement of the formula for the level-density parameter
at saddle point, Eq.~\eqref{eq:fita}, as follows:
\begin{equation}
  \widetilde{a}_\text{f}\left( U\right) =\frac{A}{k_{\infty }-r\,(k_{\infty }-k_{0})\exp
    \left( -f \frac{\kappa }{k_{\infty }-k_{0}}\frac{U}{A}\right) }\text.\label{eq:fita_r}
\end{equation}
The $r$ variable, which replaces the $a_\text{f}/a_\text{n}$ ratio, is a free parameter that describes 
the difference in the effect of long-range correlations for the saddle point.
In the limit of zero excitation energy, Eq.~\eqref{eq:fita_r} leads to
\[
\frac{a_\text{f}}{a_\text{n}}=\frac{k_0}{k_\infty-r\,(k_\infty-k_{0})}\text.
\]
while for $U\rightarrow\infty$, $a_\text{f}=a_\text{n}=A/k_{\infty}$. 
The value of $r$ thus determines the $a_\text{f}/a_\text{n}$ ratio at low energy. We expect on physical grounds that $r$
should be slightly larger than one, to reflect the increase in surface area and an enhanced 
collective enhancement of the saddle-point configuration. 
This would also lead to $a_\text{f}/a_\text{n}>1$ at small $U$. The parameter $f$, on the
other hand, expresses the different fade-out rate of long-range correlations at the saddle point compared 
to the ground state. This quantity is essentially unconstrained by experimental data. However, we observe that,
from Sec.~\ref{sec:fprob}, the approximation of an
energy-independent $a_\text{f}/a_\text{n}$ ratio is a good one at low excitation energies, since we can successfully
reproduce fission cross sections in fusion and spallation. We impose this condition by requiring that
\[
{\left.\frac{\partial(a_\text{f}/a_\text{n})}{\partial U}\right|}_{\mathrlap{U=0}}=0\text.
\]
This introduces a correlation between the parameters $f$ and $r$:
\begin{equation}
  f=\frac{k_\infty-r\,(k_\infty-k_{0})}{r\,k_0}\text.\label{eq:correl_f_r}
\end{equation}
There is no \emph{a priori} reason to expect that the fade-out rate at the saddle point (described by $f$) should be correlated
with the $a_\text{f}/a_\text{n}$ ratio at low energy. We make this assumption on a phenomenological basis.
Note that Eq.~\eqref{eq:correl_f_r} implies that $f<1$ for $r>1$, i.e., that long-range correlations
should fade out more
slowly at the saddle point than in the ground state. One should also note that $\kappa$ has a very strong mass dependence
\cite{Charity10} and therefore the modification of $a_\text{f}/a_\text{n}$ with excitation energy is much stronger for 
the \textit{p}+U reaction compared to the lighter systems. One can indeed see from Fig.~\ref{fg:tishchenko_paper1}
that this is the system that requires the biggest modification from our previous solution.

\begin{figure}[tbp]
  \includegraphics[width=\linewidth]{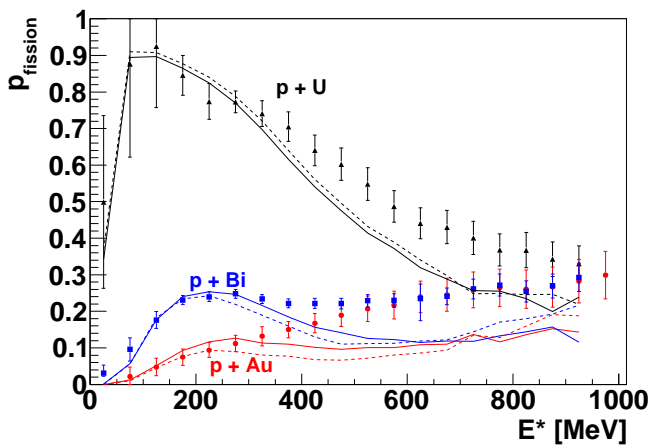}
  \caption{(Color online) Same as Fig.~\ref{fg:tishchenko_paper1}, for a Bohr-Wheeler width (solid lines) or a Lestone
    width with a 1-zs fission delay (dashed lines) with an energy-dependent effective $a_\text{f}/a_\text{n}$ ratio, Eq.~\eqref{eq:fita_r}.}
  \label{fg:tishchenko_paper2}
\end{figure}

We can finally determine the value of the $r$ parameter by requiring, for example, that the fission cross section for
1-GeV \textit{p}+$^{208}$Pb be correctly reproduced. For a Bohr-Wheeler width without fission delay, this condition
yields $r=1.0747$, which corresponds to $a_\text{f}/a_\text{n}=1.051$ for $U=0$. For a Lestone width with a 1-zs fission
delay, we get $r=1.1$ and $a_\text{f}/a_\text{n}=1.069$ at $U=0$. Fusion-fission and spallation-fission are
not severely affected by this modification, as shown by the dotted and dashed-dotted curves in Figs.~\ref{fig:fisPb}--\ref{fg:Ta_excit}. 
The resulting fission probability curves to the 2.5-GeV reactions are shown in Fig.~\ref{fg:tishchenko_paper2}. 
We have good quantitative agreement up to $\sim400$~MeV
and we can qualitatively reproduce the decrease of fission probability with excitation energy for the uranium target. This proves
that the fission probability at very high excitation energies is indeed sensitive to the fade-out of collective effects at
saddle point. 

We have thus shown that one can obtain similar quality of agreement with or without a fission delay, provided that one
increases the value of $r$.
Therefore,
while we agree with Tishchenko \etal\ \cite{Tishchenko05} that 
no transients are needed to explain their data, one cannot rule out the presence of fission transients as well.
However, the above calculations and conclusions should be taken with caution. Level densities
and fission probabilities are dramatically sensitive to all their ingredients at such high excitation energies. We have
indeed observed that the shape of the curve can also be modified by e.g.\ introducing surface terms ($\propto A^{2/3}$) in
the level-density parameter formula, or by  considering different functional forms for the fade-out
of long-range correlations. Even the competition with evaporation cannot be neglected. Finally, very high excitation energies
will eventually give rise to other phenomena, such as nuclear expansion and multifragmentation, which are not accounted
for in our framework. With all these considerations in mind, we conclude that pursuing perfect agreement between
calculations and measurements of fission probabilities for $E^*\gtrsim500$~MeV is useless for our understanding of the
physics of de-excitation.

\section{Conclusions}

We have described the first coupling of the Li\`ege Intranuclear Cascade model with the \geminipp\ compound-nucleus
de-excitation model. The fission probability was calculated using Sierk's finite-range liquid-drop fission barriers 
\cite{Sierk86} and the excitation-energy-dependent level-density parameters of Ref.~\cite{Charity10} 
adjusted to reproduce experimental kinetic-energy spectra of light particles. The latter were very important 
to obtain the correct excitation-energy dependence for heavy systems.  It was demonstrated (Sec.~\ref{sec:adjustment})
that it is possible 
to describe fission cross sections from spallation and
heavy-ion fusion reactions for $160< A < 230$ within the same framework. 
Spallation and fusion reactions populate different regions of the compound-nucleus parameter space, and thus
they probe different, but overlapping areas of the model-parameter space.
Thus, the simultaneous fitting of the statistical-decay model parameters to
spallation and fusion actually allows one to lift some of the degeneracy.  However, even with the 
large range of spin and excitation energy studied, no unique parameter set could be obtained and there remained some 
ambiguities in the choice of parameters. In particular, the effect of an increasing fission delay associated 
with fission transients could be offset by an 
increase in the parameter $a_\text{f}/a_\text{n}$, the ratio of level-density parameters at the saddle-point and ground-state 
configurations. In addition, modifications to the height of Sierk's fission barrier could be offset by scaling of the 
fission decay width which could be associated with the Kramers scaling of the Bohr-Wheeler decay width due to friction.
In spite of these ambiguities, we present two sets of statistical model parameters suitable for predictions of 
fission probabilities for spins up to 60~$\hbar$ and excitation energies up to $\sim300$~MeV.

From the study of the width of the fragment-fragment mass distributions in both fusion and spallation reactions, 
we were able to differentiate between the systematics compiled by Rusanov \etal\ based on thermal distributions 
at either the saddle or scission-point (Sec.~\ref{sec:fmass}). 
Only the saddle-point systematics provided good reproduction of the experimental data in both types of reactions, thus suggesting
the fission mass division is determined close to the saddle point. 
The asymmetry dependence of the saddle-point conditional barriers 
in the Rusanov systematics is stronger than Sierk's prediction, which produces very wide fission-fragment mass
distributions when incorporated in the Moretto formalism. Further indications could in principle be extracted from the
study of other observables, such as pre- and post-scission neutron multiplicities. However, this would happen at the
expense of introducing new parameters and ingredients for the description of the saddle-to-scission dynamics, which would
be difficult to constrain due to the lack of relevant data for spallation reactions.

We have proven (Sec.~\ref{sec:veryhigh}) that we can qualitatively describe fission probabilities at excitation energies
higher than 300~MeV by
accommodating different fade-out rates for the ground-state and the saddle-point configurations. However, we cannot
exclude that other solutions are possible, given the uncontrollable sensitivity of the predictions of the model to a large
number of its ingredients. Thus, we conclude that the theoretical uncertainties on fission probabilities at very high
excitation energy are too large to permit drawing strong conclusions about the physics of highly excited nuclei.

\begin{acknowledgments}
  The authors wish to thank S.~Leray for commenting on the manuscript, K.-H.~Schmidt for useful discussions and
  J.~Benlliure for kindly providing the experimental data for the \textit{p}+$^{181}$Ta system. This work was supported by the U.S. Department of Energy, Division of
Nuclear Physics under grant DE-FG02-87ER-40316 and by the EU IP EUROTRANS project (European Union contract number FI6W-CT-2004-516520).

\end{acknowledgments}

\end{document}